\documentclass[twocolumn,showpacs,preprintnumbers,amsmath,amssymb,amsfonts]{revtex4}

\usepackage{graphicx}
\usepackage{dcolumn}
\usepackage{bm}
\usepackage{colordvi}
\usepackage{hyperref}

\begin{document}

\preprint{APS/123-QED}

\hsize\textwidth\columnwidth\hsize \csname@twocolumnfalse\endcsname

\title{Quantitative imaging of colloidal flows}

\author{Lucio Isa$^1$}
\author{Rut Besseling$^1$}%
\author{Eric R. Weeks$^2$}%
\author{Wilson C. K. Poon$^1$}%
\affiliation{$^1$SUPA (Scottish Universities Physics Alliance) and School of Physics \& Astronomy, \\
The University of Edinburgh, Kings Buildings, Mayfield Road, Edinburgh EH9 3JZ, United Kingdom.\\  $^2$Physics
Department, Emory University, Atlanta, Georgia 30322, USA.}

\begin{abstract}
We present recent advances in the instrumentation and analysis methods for quantitative imaging of concentrated
colloidal suspensions under flow. After a brief review of colloidal imaging, we describe various flow geometries
for two and and three-dimensional (3D) imaging, including a `confocal rheoscope'. This latter combination of a
confocal microscope and a rheometer permits simultaneous characterization of rheological response and 3D
microstructural imaging. The main part of the paper discusses in detail how to identify and track particles from
confocal images taken during flow. After analyzing the performance of the most commonly used colloid tracking
algorithm by Crocker and Grier extended to flowing systems, we propose two new algorithms for reliable particle
tracking in non-uniform flows to the level of accuracy already available for quiescent systems. We illustrate the
methods by applying it to data collected from colloidal flows in three different geometries (channel flow,
parallel plate shear and cone-plate rheometry).

\end{abstract}



\maketitle

\tableofcontents

\noindent

\section{Introduction}

The last two decades have seen a surge of interest in the behavior of concentrated colloidal suspensions. These
systems have long attracted attention because of their evident practical importance. However, developments since
the 1980s have shown that well-characterized colloidal suspensions, in which the size, shape and interaction of
the particles are known, can serve as experimental model systems for understanding generic phenomena in condensed
systems. Initially, this `colloids as big atoms' approach has focussed on the use of model colloids to study
equilibrium phenomena in the bulk such as liquid structure and phase behavior
\cite{PuseyReview91,PoonReview95,poon2}. Since then, interfacial phenomena have been investigated \cite{Aarts04},
as well as bulk non-equilibrium phenomena such as phase transition kinetics \cite{AndersonReview02}, glassy arrest
\cite{Dawson2002,Sciortino05} and gelation \cite{Zaccarelli07}. In all cases, the well-characterized nature of the
experimental systems has meant that very direct comparison with theory and simulations are possible; such
synergism gives rise to rapid advances in understanding.

Most recently, the spotlight has been on the use of model colloids to study {\it driven} non-equilibrium
phenomena. In particular, coincident with intense developments in a variety of theoretical approaches
\cite{Fielding00,FuchsCates02,Schweizer05}, colloids are increasingly seen as model systems for studying the
rheology of arrested matter. Here, perhaps more so than previously, fundamental interest and immediate industrial
relevance directly coincide. Concentrated particulate suspensions, sometimes known as `pastes', have widespread
applications \cite{PasteReview06}, most (if not all) of which will involve the suspensions being mechanically
driven far away from equilibrium either as part of processing (e.g. in ceramics manufacture \cite{LewisReview00})
and/or during use. Here, as before, the study of well-characterized colloids can yield fundamental insights, many
of which are likely applicable to `real' systems with little need of `translation'. Moreover, we may expect that
driven colloidal suspensions can, in some respects, be similar to driven granular materials, themselves the focus
of intense study for both fundamental and applied reasons \cite{Jaeger1996}. Quantitative similarities of this
kind have indeed been found recently for the case of channel flow \cite{isa2}. If more such analogies are found in
the future, a unified description of colloids and grains may indeed be possible \cite{Coussot99}.

The elucidation of structure and dynamics have always been important goals in the study of colloids in general,
and of model colloids in particular. Traditionally, structural and dynamical information in this and other areas
of soft matter science is derived from scattering \cite{Bombannes}. The outputs from such experiments are the
static and dynamic structure factors. These average quantities are often directly calculable from theory, which
partly explains the appeal of scattering methods in the first place.

But the upper range of the colloidal length scale is in the optical domain, and so is amenable to direct imaging
in an optical microscope. Given the centrality of imaging in Perrin's pioneering (and Nobel Prize winning) work
using colloids to prove the existence of atoms \cite{PerrinAtoms}, it is at first sight surprising that optical
imaging has played almost no further role in the study of colloids until the last two decades of the 20th century.
But the imaging of all but the most dilute suspensions had to await two developments.

First, model systems are needed in which the refractive index of the particles can be closely, if not perfectly,
matched to that of the surrounding solvent; otherwise concentrated suspensions of large particles that are in
principle optically observable are turbid, and thus not amenable to optical imaging. A number of such systems have
been developed since the 1980s. (The development of such system also benefit the use of light scattering, which
also requires index matching.) Secondly, an imaging method needs to be found that can deal with at least a certain
degree of translucency in samples. Such a method, confocal microscopy, was invented (and patented) by Marvin
Minsky in 1955. The development of the methodology in the first few decades since its invention was driven largely
by the requirements of biologists. Since the mid-1990s, however, there has been a surge in interest in applying
confocal microscopy to the study of model concentrated colloidal suspensions. Initially, this interest was
focussed on {\em quiescent} systems \cite{weeksrev2}. In the last few years, however, it has been demonstrated
that confocal microscopy can also be used with profit to study {\em flowing} colloids, and thus yield unique
insights into the rheology of pastes. The purpose of this work is to set out in detail, for the first time, how
this can be done.

The rest of the paper is organized as follows. We first briefly review the use of imaging methods to study
colloids. We then describe in Section~\ref{sec:Instrument} new hardware that we have developed to image colloidal
flows in various geometries. In Section~\ref{sec:Locating} we review the basics of feature identification in
(confocal) images and discuss various limits to particle identification in 2D and 3D flow fields. The core of the
work is Section \ref{sec:Algorithm}, where we turn to particle tracking. In Section~\ref{subsec:evaluation} we
evaluate for the first time, using data from simulations, the applicability of a classic (and widely applied)
tracking algorithm \cite{CrockerGrierJColIntSc96_tracking} in quiescent and sheared systems where the average
motion is zero. We then describe in detail, Sections.~\ref{subsec:iterated} and \ref{subsec:cit}, our new methods
to track particles in the presence of flow. Finally, in Section~\ref{sec:Examp} we demonstrate the applicability
of these methods to imaging colloidal flows in various geometries, including `home built' environments and a
commercial rheometer.

\section{Imaging colloidal suspensions}

The imaging of a single layer of colloids is unproblematic, and has been used to great effect to study fundamental
processes in 2D, as illustrated by the work of Maret and his coworkers (e.g.~\cite{Maret00,Maret03,Maret08}). Here
we concentrate on imaging in 3D.

The use of conventional (non-confocal) optical microscopy to study concentrated colloidal suspensions in 3D has
been reviewed before \cite{Elliot01}. In nearly index-matched suspensions, contrast is generated using either
phase contrast or differential interference contrast (DIC) techniques. One advantage of conventional microscopy is
speed: image frames can easily be acquired at video rate. But it has poor `optical sectioning' due to the presence
of significant out-of-focus information, so that particle coordinates in concentrated systems cannot be
reconstructed in general, although structural information is still obtainable under special circumstances
\cite{Bristol97}.

Compared to conventional microscopy, confocal microscopy delivers superior `optical sectioning' by using a pinhole
in a plane conjugate with the focal ($xy$) plane. It allows a crisp 3D image to be built up from a stack of 2D
images. But each 2D image needs to be acquired by scanning, which imposes limits on its speed. The technique has
been described in detail before \cite{wilson}.

The use of confocal microscopy in the study of concentrated colloidal suspensions was pioneered by van Blaaderen
and Wiltzius \cite{vanblaaderen1}, who show that the structure of a random close packed sediment could be
reconstructed at the single particle level. The confocal microscopy of colloidal suspensions in the absence of
flow has been reviewed recently \cite{dinsmore,habdas02,weeksrev1,weeksrev2}, and we refer the reader to these
reviews for details and references. Here, we simply note that this methodology gives direct access to {\em local}
processes, such as crystal nucleation \cite{Weitz01} and dynamic heterogeneities in hard-sphere suspensions near
the glass transition \cite{kegel,weeks1}.

In this work, we focus on the use of confocal microscopy for imaging colloids {\em under flow}, or confocal
`rheo-imaging' (reviewed in \cite{PoonReview09}). Conventional rheology studies the mechanical response of bulk
samples. As far as the study of concentrated, model suspensions is concerned, much attention has been given in the
last few years to {\em non-linear} rheological phenomena, e.g., the different ways in which repulsion- and
attraction-dominated colloidal glasses yield \cite{pham3,Pham08}. The bulk rheological data are consistent with
the former yielding by a single-step process of `cage breaking', and the latter yielding in two steps, first
breaking interparticle bonds, and then breaking nearest-neighbor cages. Confocal imaging can play a decisive role
in the verification of such microscopic interpretation, which inevitably makes reference to local processes on the
single-particle level.

Moreover, direct imaging can clearly shed light on complicating factors in conventional rheological measurements
such as wall slip \cite{RusselGrantColSurfA2000_slipyield,BuscalJRheo93_slip} and flow non-uniformities such as
shear banding \cite{Olmsted08}. Here, significant progress can be made without imaging at single-particle
resolution, by using various coarse-grained velocimetry methods. Traditional Particle Image Velocimetry (PIV)
\cite{PIV} requires transparent samples. This technique has recently been used in a rheometer to give important
information on slip in emulsions \cite{MeekerPRL04_rheoslip,MeekerJRheo2004_slipandflowrheo}. Other methods for
velocimetry have been developed that has no requirement for transparency, such as heterodyne light-scattering
\cite{SalmonEPJAP2003_heterodyneDLS} and ultrasonic velocimetry \cite{MannevilleEPJAP04_ultrasound}. The latter
has been applied to characterize slip and flow nonlinearities in micelles and emulsions
\cite{BecuPRL06_twoemulsions,BecuPRE07_rheomic}.

A robust method for velocimetry which can also provide additional information
on the density profile is Nuclear Magnetic Resonance Imaging (NMRI)
\cite{fukushima_NMRAnnRevFlMech99,CallaghanRepProgPhys1999_rheoNMR,BonnAnnRevFlMech2008_NMR,GladdenMeasSciTech96_NMR}.
The technique has spatial resolution down to $\sim
20$~$\mu$m and has been combined with rheometric
setups to relate velocity profiles to macroscopic
rheology~\cite{ovarlez2,RaynaudCoussotJRheo02_NMRyield}. This
approach has been used to investigate the occurrence of shear
bands~\cite{HuangBonnPRL04_wetgranularflow} and shear
thickening~\cite{fall}.

Thus, both PIV and NMRI give additional insight unavailable from bulk rheology alone. But to build up a complete
picture of colloidal flow, it is desirable also to have information on the single particle level. For this
purpose, a method related to PIV and particle tracking has been applied to non-Brownian suspensions and allowed
the measurement of non-affine particle motion and diffusivity
\cite{BreedveldJFlMech98,BreedveldPRE2001_nonbrowniantracerdiffusion}. But {\em direct} imaging of the
microstructure during flow is needed to give complete microstructural information. Optical microscopy has this
capability.

It is possible to use conventional (non-confocal) video microscopy to study shear effects in 3D
\cite{haw2,haw3,SmithPoonPRE2007,BiehlPalbergEPL04}. But the poor optical sectioning hinders complete,
quantitative image analysis. Confocal microscopy significantly improves sectioning, and permits in principle the
extraction of particle coordinates. But the need for scanning initially meant rather slow data acquisition rates,
so that observations in real time (i.e. during shear) produced blurred images that again limited the potential for
quantitative analysis \cite{tolpekin}. A common solution was to apply shear, and then image immediately after the
cessation of shear, both in 2D~\cite{hoekstra,stancik} and in
3D~\cite{VaradanSolomonJRheo2003_gelflowconfocal,tolpekin,CohenWeitzPRL04_confinedshear,SolomonJCP06_shearcrystalCF}.
(Earlier work using conventional video microscopy~\cite{haw2,haw3} resorted to the same strategy.)

More recently, the availability of fast confocal systems (see Section~\ref{subsec:scanning}) means that
nearly-real-time reconstruction of structure during flow in 3D at single-particle resolution has become possible.
Such experiments face two key challenges: sample environment and data analysis. First, the flow geometry used
clearly must be compatible with the optical requirements of simultaneous confocal imaging. A number of different
arrangements have been demonstrated to date. Derks and co-workers carried out a first experimental study by using
a counter-rotating cone and plate shear cell combined with a fast confocal microscope~\cite{derks} and obtained
particle coordinates and tracks in the zero-velocity plane as well as velocity profiles across the geometry gap.
The same group has recently produced a more sophisticated set up which uses a parallel plate shear cell~\cite{wu2}
capable of spanning a vast range of shear rates and frequencies which they used to study crystallization of
colloids under shear. A parallel plate shear cell has also been used by Besseling and
co-workers~\cite{BesselingPRL2007} to study the shear-induced relaxation in hard-sphere colloidal glasses, while
recent experiments by Isa and co-workers~\cite{isa2} have elucidated the behaviour of colloidal sediments flowing
into micro-channels. In this work, we give the details for two of these geometries \cite{BesselingPRL2007,isa2},
and describe and demonstrate a new one: the coupling of a fast confocal scanner to a commercial rheometer, which
allows simultaneous confocal imaging and full rheological characterization of the same sample.

The second challenge is data analysis: how to extract accurate particle coordinates from raw image stacks. In
particular, special methods are needed for reliable tracking, since the large displacements from frame to frame
imposed by flow may inhibit correct identification of particles between frames. The same problem confronts the use
of imaging to study granular flows \cite{XuReeveseRSciInstr04_trackingerrors}. In this work, we describe in detail
a new method for tracking particles from confocal images acquired during flow. We demonstrate its correctness and
measure its limitations by using data from computer simulations, as well as illustrate its use with real
experimental data.

\section{Materials and Instrumentation}
\label{sec:Instrument}

\subsection{The colloidal particles}

Our goal is to perform confocal imaging in real time at single-particle resolution of colloidal suspensions under
flow. Confocal microscopy is, in principle, able to imaging `inside' slightly turbid systems, but the image
quality deteriorates with sample turbidity. In order to obtain as sharp images as possible to test the limits of
our methodology, we performed experiments using a index-matched suspension that is optically clear.

The particles were poly--methyl--methacrylate (PMMA) spheres, sterically-stabilized by chemically-grafting
poly--12--hydroxy stearic acid (PHSA)~\cite{antl}. The particles can be dyed with a fluorophore (NBD, 4 chloro--7
nitrobenz--2 oxa 1,3 diazole), which is excited at 488 nm and emits at 525 nm. Particles can be suspended in a
mixture of decalin (mixed-decahydronaphtalene, Sigma--Aldrich, $n_{\rm decalin} = 1.4725 \pm 0.0005$) and tetralin
(tetrahydronaphtalene, Sigma--Aldrich, $n_{\rm tetralin} \simeq 1.5410 \pm 0.0005$) to achieve full refractive
index matching of the solvent and the particles ($n_{\rm susp} \simeq 1.5$). This matching assures hard-spheres
interactions~\cite{ackerson} and also limits scattering of both the laser and the excited light during confocal
microscopy. However, the decalin-tetralin mixture has a lower density than PMMA ($1.188\rm{g/cm}^3$). To achieve
buoyancy-matching, particles can be suspended in a mixture of cyclo-heptyl-bromide (CHB) and
mixed-decalin~\cite{yethiraj}. The buoyancy-matching composition also closely matches the refractive index of the
suspension ($n_{\rm susp} = 1.494$) \cite{dinsmore}. The addition of CHB to a hard-spheres suspension induces
charge on the particles, which can be screened by adding a suitable amount of salt~\cite{yethiraj}
(tetrabutylammonium chloride, $(\mathrm{C_4 H _9)_4NCl}$, $\mathrm{M_W = 277.92 g/mol}$, Fluka).

The buoyancy-matching is very sensitive to temperature changes; the thermal expansion coefficient of the solvent
exceeds that of PMMA by about a factor ten and a decalin-CHB mixture of a given composition will therefore match
the particle density only in a very narrow temperature range~\cite{lu2}. We exploit this fact to prepare
suspensions of different volume fractions by centrifuging the suspension at a temperature higher than the
buoyancy-matching one. Finally, imaging can either be performed
on a fully fluorescent sample or on refractive index matched
systems seeded with fluorescent particles. In the course of our description we shall specify the details of the
system used in each example.

\subsection{The confocal microscope} \label{subsec:scanning}

To perform confocal imaging during flow, high acquisition rates and thus fast laser scanning methods are required.
Among these are spinning disk systems \cite{wilson} (with possible micro-lens array extension) or laser scanning
by resonant galvanometric mirrors. The confocal scanner we use (VT-Eye, Visitech International, with a solid state
$488$~nm laser) employs a combination of a standard galvanometer and an acousto-optic deflector (AOD)~\footnote{To
accommodate the wavelength dependent deflection of the AOD, the instrument uses a slit instead of a pinhole, but
in practice the resolution is very similar to that of standard pinhole configuration.}. The former positions the
laser beam at a certain $y$-position, while the AOD much more rapidly scans a line along $x$. The acquisition rate
is thus mainly determined by the `slow' galvanometer. Typical frame rates for 2D image series range from $f_{\rm
scan}=5$~Hz  for images of $1024 \times 1024$ pixels to $f_{\rm scan}=100-200$~Hz for images of $256 \times 256$
pixels. The upper limits on colloid diffusivity or flow speed imposed by these acquisition rates are described in
Sec.~\ref{subsec:loclimit}.

\begin{figure}
\includegraphics[width=0.45\textwidth,clip]{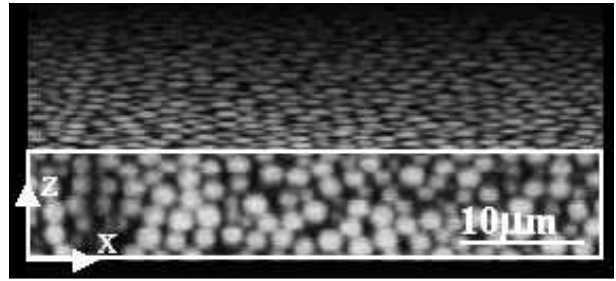}
\caption{Projection of a raw 3D image stack of a concentrated colloidal suspension, acquired in $\sim 1$~s. The
thickness is $10~\mu$m.}\label{fig:1}
\end{figure}

We have imaged flow in a parallel-plate shear cell and in square capillaries using the `standard configuration',
where the confocal scanner is coupled to a Nikon TE Eclipse $300$ inverted microscope, with a $100 \times$ or $60
\times$ magnification, oil-immersion objective with a numerical aperture ($NA$) of $1.4$. The depth of the focal
plane, $z$, is controlled by a piezo-element mounted on the microscope nosepiece. For 3D imaging, a $z$-stack of
2D images is collected (Fig.~\ref{fig:1}) by rapid variation of the height of the piezo and synchronized 2D
acquisition at each $z$. The corresponding 3D acquisition time is $N_z/f_{\rm scan}$ with $N_z$ the number of 2D
slices. We have also coupled the confocal scanner to a commercial rheometer, to enable simultaneous imaging and
rheological measurements on the same sample.

\subsection{Shear cell}

\begin{figure}
\includegraphics[width=0.45\textwidth,clip]{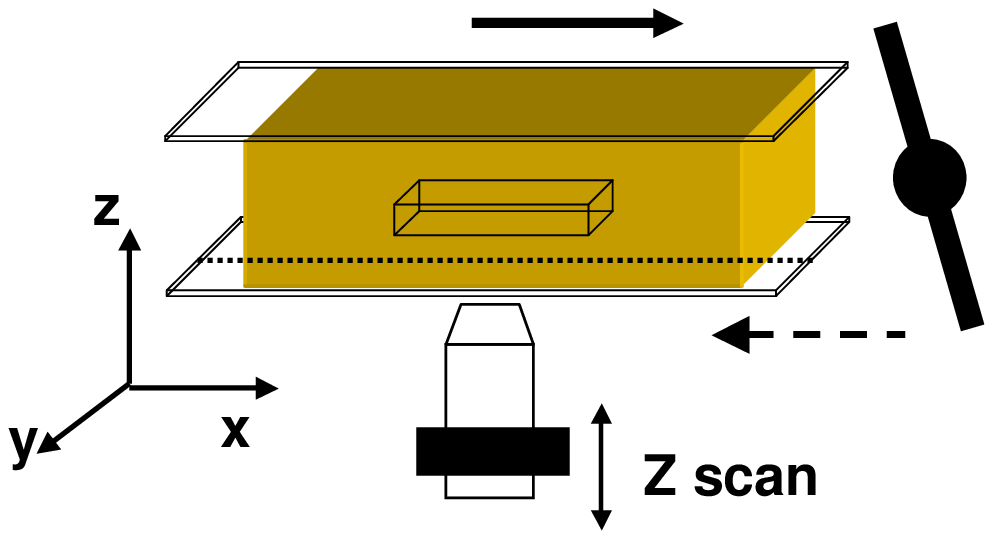}
\caption{Sketch of the shear cell. The sample is positioned between two parallel slides; the top one is driven by
a mechanical actuator while the bottom one can either be fixed or be translated in the opposite direction (dashed
arrow). The suspension is imaged from below (image volume highlighted) with a confocal microscope whose focal
depth is controlled by a piezo-electric motor.}\label{fig:2}
\end{figure}

\begin{figure}
\includegraphics[width=0.45\textwidth,clip]{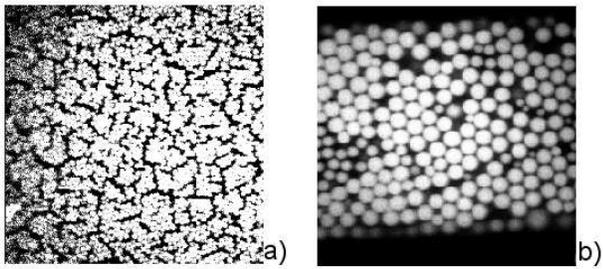}
\caption{Confocal images of the coating (a) on a cover slide for shear flow imaging (image size
$56~\times~56~\mu{\rm m}^2$), (b) on the inner surface of a $50~\times~50~\mu{\rm m}^2$ glass capillary (image
size $43~\times~43~\mu{\rm m}^2$). The larger particles in (b) form the coating, the smaller ones are suspended
and flowing.}\label{fig:3}
\end{figure}

Initial experiments on shear flow were performed with a linear
parallel-plate shear cell (plate separation $Z_{gap}\sim 400-800$
$\mu$m, parallel to $\pm 5$~$\mu$m over a lateral distance of
$2$~cm) where the top plate is driven at $0.05-10$ ~$\mu$m/s by a
mechanical actuator with magnetic encoder. We denote the velocity,
vorticity and velocity gradient direction by $x$, $y$, and $z$,
respectively, as shown in Fig.~\ref{fig:2}. The
maximum relative plate translation along $x$ is $L_s \sim 1$~cm,
so that steady shear can be applied up to a total accumulated
strain $\Delta \gamma=L_s/Z_{gap} \gtrsim 1000 \%$. The cell can
be operated either with the bottom plate fixed or with the plates
counter propagating via an adjustable lever system, which allows the height of
the zero velocity plane to be set at any distance from the bottom
plate. A drop of suspension (covering an area of $\sim 200$~mm$^2$) is confined between the
plates by surface tension. A solvent bath surrounding the plates
minimizes evaporation. Wall slip, prominent in glassy systems
\cite{Ballesta2008}, and wall-induced ordering were prevented
by sintering a concentrated, disordered layer of particles -- obtained by
spincoating a $\phi \sim 30\%$ suspension -- onto the glass surfaces,
Figure~\ref{fig:3}(a). We typically image a $30 $ $\mu$m
$\times$ $30$ $\mu$m $\times$ $15 $ $\mu$m volume in the drop (with
$\sim 3000$ particles), up to $\sim 40$~$\mu$m above the coverslide.

\subsection{Capillary flow}

\begin{figure}
\includegraphics[width=0.45\textwidth,clip]{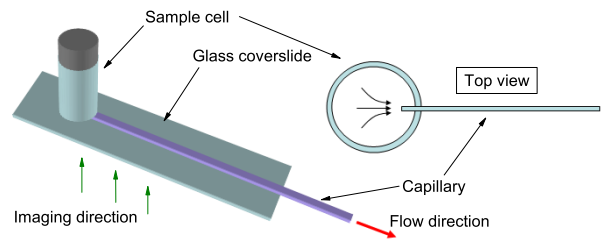}
\caption{Sketch of the sample cell for capillary flow. The capillary is not drawn to scale. The construction is
placed on the microscope stage plate and the flow is imaged from below.}\label{fig:4}
\end{figure}

We have also studied the flow of concentrated colloidal suspensions in glass micro-channels~\cite{isa1,isa2}.
Figure~\ref{fig:4} shows a sketch of a sample cell for such experiments. It is assembled by gluing a glass
capillary onto a microscope coverslide with UV curing glue (Norland Optical Adhesive) by exposing it to UV light
for a few minutes. Once the glass channel is attached, a glass vial (1.5 cm diameter), the bottom of which is
removed, is glued on top of one end of the channel also with UV glue (2 hours exposure). Finally a PVC tube
($1$~mm internal diameter) is connected to the free end of the capillary and the connection is sealed with epoxy
glue. The sample cell is positioned onto the microscope stage with the cover slide, forming the bottom of the
cell, in direct contact with the microscope objective via the immersion oil.

All channels are borosilicate glass capillaries (Vitrocom) with rectangular ($20 \times 200 \mu{\rm m}^2$, $30
\times 300 \mu{\rm m}^2$, $40 \times 400 \mu{\rm m}^2$) or square cross sections ($50 \times 50 \mu{\rm m}^2$, $80
\times 80 \mu$m, $100 \times 100 \mu$m) and a length of 10 cm. The capillaries can either be used untreated,
i.e.~smooth on the particle scale, or have their inner walls coated with a sintered disordered layer of PMMA
particles to ensure rough boundaries (Figure~\ref{fig:3} b). This is achieved by filling the capillary with a
15--20\% suspension of similar particles and subsequent drying in a vacuum oven at 110-120 $^{\circ}$C.

\begin{figure}
\includegraphics[width=0.45\textwidth,clip]{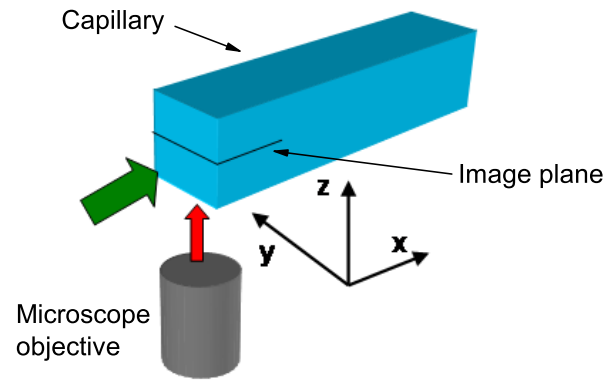}
\caption{Close up of the experimental geometry. For clarity only the capillary and objective are drawn. The flow
(green arrow) is imaged from below and 2D $xy$ slices are collected at a depth $z$.}\label{fig:5}
\end{figure}

The imaging geometry is sketched in Fig.~\ref{fig:5}. The flow is along the channel in the $x$
direction; by adjusting the image size we can capture the entire cross section of the capillary. Using a modified
microscope stage plate with a long rectangular slot instead of the standard circular aperture, the imaging can be
performed at different positions along the channel, over a range of $~\lesssim 5$~cm. The suspension is first
loaded into the sample cell and then driven along the channel by a constant pressure difference achieved by
motorized displacement of a syringe plunger connected to the PVC tubing. During imaging, the pressure is monitored
with a pressure gauge (MKS Series 902 Piezo Transducer).

\subsection{Confocal rheoscope}
\label{subsec:cfrheoscope}

\begin{figure}
\includegraphics[width=0.3\textwidth,angle=-90,clip]{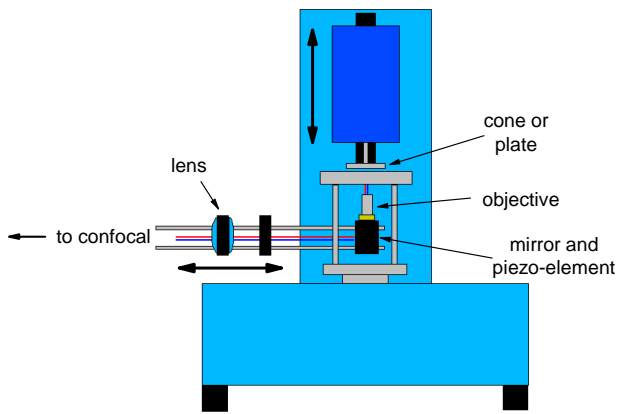}
\caption{Schematics of the confocal rheoscope. The top arrow marks translation of the rheometer head to adjust the
geometry gap, the horizontal arrow indicates translation of the arm supporting the objective to image at different
radial positions.}\label{fig:6}
\end{figure}

\begin{figure}
\includegraphics[width=0.45\textwidth,clip]{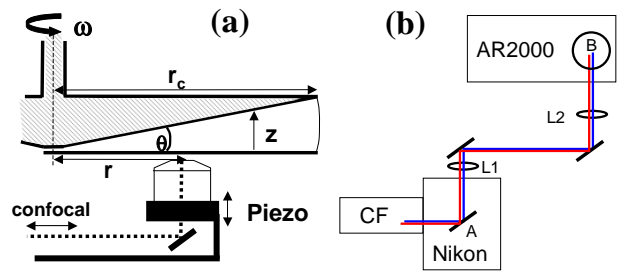}
\caption{(a) Zoom in on the central part of the confocal rheoscope. The position of the mirror, directly
underneath the piezo, is indicated as {\bf B} in the right figure. (b) Global sketch of our implementation. The
optical path (colored) from the confocal (CF) is guided from the back of the microscope (Nikon) via mirrors and
two lenses $L1$ and $L2$, to position {\bf B} corresponding to the mirror in (a). The distance between lenses $L1$
and $L2$ is $2f+d$, with $f$ the focal length. The distance between the standard position of the objective back
aperture (above {\bf A}) and the new one (above {\bf B}) is $4f+d$, where $\mid d \mid \lesssim 1$~cm allows for
lateral positioning underneath the plate.}\label{fig:7}
\end{figure}

In order to perform confocal imaging of the flow and simultaneously obtain the global suspension rheology, we have
combined the fast confocal scanner with a stress controlled rheometer (AR2000, TA Instruments), Figs.~\ref{fig:6}
and ~\ref{fig:7}. The rheometer has a custom-built, open base construction, mounted on its normal force sensor,
with pillars providing space for a mirror and objective mounted on a piezo-element, which makes it possible to
vary the depth of the focal plane. An aluminium plate with imaging slit is mounted on the pillars, and can be
accurately leveled via three adjustment screws. A glass slide (radius $2.5$~cm, thickness $\sim 180$~$\mu$m),
mounted on the plate, forms the bottom surface of the measurement geometry through which the imaging is performed.
The rheometer can thus be operated in plate-plate or cone-plate geometry, but generally we used a stainless steel
cone of radius $r_c=20$~mm and cone angle $\theta=1^{\circ}$. Both the glass slide and the cone can be made rough
on the particle scale using the spincoating and sintering method. Evaporation can be minimized by a solvent trap,
but we found superior reproducibility of the rheology of the most concentrated suspensions by slightly
under-loading the geometry and applying a tiny rim of immiscible liquid (glycerol) around the geometry edge.
Finally, we checked that bending of the cover slide was negligible, see  Sec.~\ref{subsec:rheoscopeapp}.

The confocal scanner is coupled to the optics under the plate. In our setup, the scanner remains fixed in the
standard configuration, connected via a C-mount to the Nikon microscope, see Fig.~\ref{fig:7}(b). To
provide the coupling, we altered the optical path of the laser and the excited light. By positioning a movable
mirror, the beam exits through the rear of the microscope (Fig.~\ref{fig:7}(b)) and then passes through
additional mirrors and two lenses; one of the lenses is mounted on a mechanical `arm', at the end of which,
situated below the transparent plate, a final mirror and the piezo objective mount are located. The two lenses
provide one-to-one imaging of the back-aperture of the objective in its standard position on the nosepiece of the
Nikon, onto that of the objective in its new position under the transparent rheometer plate. During imaging, the
piezo-element is controlled by the software of the confocal microscope, providing the same 3D imaging capability
as in normal operation.

\section{Particle location and its limitations}
\label{sec:Locating}

\subsection{Locating and identifying the particles}
\label{subsec:loc}

The first step to obtain quantitative information on particle dynamics from the images is to locate the particles.
The most widely used algorithm for this purpose to date in colloid science is
that of Crocker and Grier
(CG)~\cite{CrockerGrierJColIntSc96_tracking}, with relevant software
 in the public domain \cite{idlwebsite}.

Three main assumptions are needed in order to locate and identify the particles. The features must appear as
bright objects onto a dark background, we assume that they are spherical in shape~\footnote{In actual fact, due to
a possible different pixel size in the $x$, $y$ and $z$ directions the images may not appear as spherical; the
crucial assumption is that the imaged features are \textit{spherical in reality}. Any stretching of the image can
then be removed prior to the location procedure.}  , and that the maximum in the brightness of a feature
corresponds to its center. The concepts at the basis of feature location are still applicable to objects which do
not follow these requirements but the practical algorithm for locating them will be different and generally more
complicated, see e.g.~\cite{BrangwynneKoenderinkBioPhysJ2007_filamenttracking}.

Since the particle centers are identified in terms of their
intensity, undesired intensity modulations which can give rise
to mistakes in particle location need to be eliminated. This is
achieved via image filtering using a spatial bandpass filter.
This eliminates any long wavelength contrast gradients and also
short wavelength pixel to pixel noise.

The coordinates of the centers of the features are initially obtained by locating the local intensity maxima in
the filtered images. A pixel corresponds to a particle center if no other pixel has a higher intensity within a
given distance to it; typically this distance is slightly larger than the average particle radius. These
coordinates are then refined to get the positions of the particle centers with a higher accuracy by applying a
\textit{centroiding} algorithm which locates the brightness-weighted center of mass (centroid) of the particles.
With this refinement procedure the coordinates of the particle centers can be obtained with sub pixel resolution
down to less than 1/10 of the pixel size~\cite{CrockerGrierJColIntSc96_tracking}. The centroiding procedure proves
itself effective but has some limitations. Correctly locating the particles becomes more difficult as the system
becomes more concentrated and individual particle images may start to overlap. This difficulty can be dealt with
by fluorescent labeling only the particle cores \cite{kegel} so that images are well separated even at the highest
densities. When such particles are not available, improvements in coordinate refinement may be required. Such an
improvement, based on fitting the intensity profile of the particle to the `sphere spread function' (SSF), has
been devised by Jenkins ~\cite{JenkinsEgelhaafAdvColIntSci08}. We have used this method successfully for our 3D
images described in Sec.~\ref{subsec:3Dtrack}. Finally, to avoid edge effects, particle centers identified within
a radius from the image edge are ignored.

As noted above, the coordinates of particle centers are often found to an accuracy tied to the pixel size.  This
correctly implies that modifying the microscope optics so that the size of a pixel is smaller will improve the
location of particle centers.  In general, if the image of a particle is $N$ pixels across, the center of that
particle can be found to an accuracy of (pixel size)$/N$. It is important to recognize that this accuracy is
different from, and often better than, the optical resolution of the microscope. The optical resolution relates to
telling the difference between two closely positioned bright objects: if they are closer together than the
resolution, then their diffraction-limited images blur together in the image.  The resolution limit for an optical
microscope is given by the wavelength of light used and the numerical aperture (NA) of the objective lens, as
$\lambda / (2 {\rm NA})$, and the best resolution for an optical microscope is about 0.2~$\mu$m. This figure
reflects the wave nature of light. The accuracy with which particle centers can be located is set by different
physical constraints, e.g. the fact that particles cannot physically overlap (although their images may)
\cite{CrockerGrierJColIntSc96_tracking}, and the knowledge that they are spherical.

\subsection{Limitations}
\label{subsec:loclimit}

The above mentioned accuracy of locating particles is intrinsic, and applies even for particles `frozen' in the
image. It can be estimated from the plateau value of the mean squared frame to frame displacements (which we
denote as MSFD) measured in a close packed sediment, where particles are essentially immobile. In general however,
additional errors on the exact center position arise from short time diffusive motion and flow advection during
acquisition of the particle image, if the images are from a scanning system (such as a confocal microscope) which
does not acquire each pixel simultaneously.

Let us estimate these effects for 2D and 3D imaging. For the 2D case, given the frame acquisition rate $f_{\rm
scan}$ and the image size in pixels ($n \times m$), the time required to scan one line is $\simeq 1/(nf_{\rm
scan})$. If $\tilde{R}$ is the particle radius in pixels, then the time required to image a particle is:
\begin{equation}
\displaystyle{t_{\rm im}^{2D}=2\tilde{R}/(nf_{\rm scan})}.
\label{eq:ti2D}
\end{equation}
With our imaging system, a $256 \times 256$ pixel image can be
taken at $f_{\rm scan}=90$~Hz. Using a $100 \times$ magnification
objective, for which the $xy$ pixel size is $\sim 0.2$~$\mu$m,
the time to acquire a 2D image of a particle with $R=1$~$\mu$m is
therefore $t_{\rm im}^{2D} \simeq 0.4$~ms.

For 3D images, acquired as a $z$-stack of 2D slices, the
limiting factor is the speed at which the particle is imaged in
the $z$-direction. The voxel size in the $z$ direction (i.e.~the
$z$-spacing between the 2D slices) may differ from that in the $x$
and $y$ direction. Denoting the particle radius in $z$-pixels by
$\tilde{R}_{\rm z}$, the time required for a 3D particle image is:
\begin{equation}
\displaystyle{t_{\rm im}^{3D} = 2\tilde{R}_{\rm z}/f_{\rm scan}},
\label{eq:ti3D}
\end{equation}
with $f_{\rm scan}$ the acquisition rate for a complete 2D image
as before. Thus, for the 3D case, using again $f_{\rm scan}=90$~Hz
and a typical value of $0.2$~$\mu$m for the $z$-pixel size, the
acquisition time for our $R=1$~$\mu$m particle is $t_{\rm im}^{3D}
\simeq 0.1$~s.

We first consider the (short time) diffusive motion of colloids in a suspension on these time scales. In the
dilute limit, the diffusion constant is $D_{{\rm s},0} = (k_{\rm B} T)/(6 \pi \eta R)$, with $k_{\rm B}$ the
Boltzmann constant and $\eta$ the solvent viscosity. For concentrated suspensions the short time diffusion
constant is reduced due to hydrodynamic hindering, $D_{\rm s}(\phi)=D_{{\rm s},0} H(\phi)$  with $H(\phi)<1$
\cite{pusey_hydro,BeenhakkerMazurPhys84,TokuyamaPRE94_diffusion,BradyJChemPhys93_rheocol,MegenPRE98_tracersinglass}.
The average motion in one direction during the acquisition time is $\sqrt{2D_{\rm s}(\phi) t_{\rm im}}$, i.e.~the
additional error is $\delta = \sqrt{D_{\rm s}(\phi) t_{\rm im}/2}$. Using $\eta = 2.7 \times 10^{-3}$~Pa$\cdot$s
for our solvent (decalin) and $T=300$~K, $D_{{\rm s},0}= 8.13\times 10^{-2} \mu\rm{m}^2/\rm{s}$ for the
$R=1$~$\mu$m colloid. The resulting error due to thermal displacement during 2D imaging of a dilute system is then
$\delta_{2D} \simeq 2$~nm, while for the 3D case, using the same frame rate and $z$-pixel size as above, we have
$\delta_{3D} \simeq 35$~nm. While the former is considerably smaller than the intrinsic 1/10 pixel accuracy ($\sim
20$~nm), for 3D the thermal motion is the limiting factor. Note that these considerations apply to hard sphere
systems only; when additional interactions limit the short time displacements, the intrinsic $1/10$ pixel limit
may still apply.

Flow can also induce additional errors on the particle location due to image distortion. Since the image of a
particle is scanned either via lines in 2D or via horizontal slices in 3D, it will be distorted because the
particle is displaced between two consecutive lines as well as `slices'. Such distortion can be exploited to
deduce the local flow velocity, see~\cite{derks}, but here we are interested in the velocity range for which the
distortion is sufficiently small to consider the object as effectively spherical. The particle speed beyond which
this no longer holds can be estimated by comparing the imaging time $t_{\rm im}$ with the time $t_{\rm f}$
required for the flow to displace the particle over its own diameter. For a flow velocity $\tilde{V}$ (in pixels
per second), $t_{\rm f}=2\tilde{R}/\tilde{V}$. We consider the particle significantly distorted if $t_{\rm
im}/t_{\rm f} \ge 0.1$, i.e.~a distortion of $1$ pixel for a particle size $2\tilde{R}=10$. Using
Eqs.~\ref{eq:ti2D},\ref{eq:ti3D} for the acquisition times, the maximum velocities are:
\begin{equation}
\displaystyle{\tilde{V}^{max}_{2D} = 0.1 n f_{\rm scan},~~~~~~~~~\tilde{V}^{max}_{3D}=0.1 f \tilde{R}/\tilde{R}_{\rm z}}.
\label{eq:distortionlimit}
\end{equation}
For typical parameters ($f_{\rm scan}=90$~Hz, $n=256$~pixels, 1 pixel
$\simeq 0.2$~$\mu$m) the limiting velocity in 2D is $V^{max}
\simeq 500$~$\mu$m/s, while for 3D images we obtain $V^{max}
\sim 2~$~$\mu$m/s for an $x$ to $z$ pixel size ratio of 1:1. In
both cases, further improvement could be achieved by removing the
distortion prior to locating the particles via image correlation
procedures~\cite{derks} or by quantitatively accounting for
distortions and performing the centroiding algorithm using a
`distorted particle template'.

\section{Tracking algorithms}
\label{sec:Algorithm}

Once the coordinates have been found in each frame, they need to be merged into trajectories describing the
particle motion. In this `tracking' procedure, each particle is labeled with an identification tag and an
algorithm looks for particles in the following frame that can be assigned the same tag. Tracking has applications
in fields as diverse as robotics and biophysics
\cite{SteagerAPL07_baterialtrack,BrangwynneKoenderinkBioPhysJ2007_filamenttracking,RogersBioPhJ2008_tracking}. In
the field of digital image processing a variety of methods has been devised \cite{chetverikov01particle}. In each
method, a specific cost function is calculated based on the changes in coordinates for each set of
identifications, possibly extended with a cost function for change in feature appearance
\cite{chetverikov01particle}. The `correct' identification is then obtained as the one for which the cost function
is minimized.

\subsection{The classic CG algorithm}
\label{subsec:evaluation}

\subsubsection{Tracking algorithm}
\label{subsubsec:tracking}

The algorithm we use, devised by Crocker and Grier (CG), is
based on the dynamic properties of non-interacting colloids
\cite{CrockerGrierJColIntSc96_tracking}. The cost function in
this case is the mean squared frame to frame displacement (MSFD,
as defined before) of particles between frames. Given the position
of a particle in a frame and all the possible new positions, in
the following frame, within a `tracking range' $R_T$ of the old
position, the algorithm chooses the identification which results
in the minimum MSFD.  Particles moving farther than $R_T$ between
frames are unable to be tracked, and are either mis-identified as
other particles, or else treated by the algorithm as new particles.

Note that the original CG algorithm also includes the ideas
discussed in Sec.~\ref{sec:Locating}; here we focus on tracking
particles between successive frames, rather than locating
particles in a single image.  These two parts of the CG algorithm
are decoupled: the tracking method works independently of
how the particles were originally identified.

\subsubsection{Hard particle simulations}
\label{subsubsec:simulations}

Crocker and Grier tested their algorithm by tracking the self diffusion of particles in dilute colloidal
suspensions~\cite{CrockerGrierJColIntSc96_tracking}. Since then, the algorithm has been applied in various studies
of quiescent colloidal systems~\cite{weeksrev1,weeksrev2}. However, quantitative studies of its tracking
performance in realistic concentrated systems, possibly with additional motion on top of Brownian diffusion, have
not been performed to date. In fact, the study of concentrated systems explicitly pushes the CG tracking algorithm
beyond its design parameters.  We therefore apply the classic CG algorithm to computer generated data, in which
the particle identity is known {\it a priori}, and evaluate its performance for quiescent systems of different
densities and for various imposed particle motions. Quiescent data were generated by Monte-Carlo (MC) simulation
of hard-disks in two dimensions and of hard-spheres in 3D (see e.g.~\cite{DoliwaPRL98_HSglasssimu_hetero})
imposing mean squared displacements between each MC iteration chosen to obtain a sufficient success rate. We
simulated $N>1200$ particles and, as in experimental data, particles may (dis)appear at arbitrary times around the
edge of our simulation cell. At sufficiently high concentration, the dependence of the mean squared displacement
on the lag time, $\langle dr^2(\tau) \rangle$ (the `MSD'), shows a transition from short (`in-cage') to long-time
(`cage breaking') diffusion, as expected for dense fluids \cite{DoliwaPRL98_HSglasssimu_hetero} \footnote{In 2D,
the hard-disk system shows dislocation mediated, two stage, melting according to the Kosterlitz-Thouless scenario.
We find a melting density $\phi_{2D} \sim 0.7$ \cite{Jaster}. For 3D our densities are also below the melting
fraction $\phi_{3D}^{\rm F} = 0.494$. Our simulations are by no means exhaustive, they are merely performed to
test the success of the tracking algorithm.}.

We take the data from the simulations, and treat it as the raw
data of particle positions for the classic CG-algorithm.  Specifically,
we use the CG-algorithm to track particles between MC-iterations
$i$ and $i+n$, for which the {\it true} MSFD is $\langle \sum_j
(r^j_{i+n}-r^j_i)^2 \rangle = \langle dr^2_n \rangle$, with $j =
x,y,z$.  The key idea of the CG-algorithm is that ideally, between
each frame of the movie, the majority of particles should move less
than the typical interparticle (center-to-center) spacing $\Delta$.  In other words,
it is desirable for the MSFD to be less than $\Delta^2$. In fact,
since the MSFD is calculated by averaging over {\em all} particles, many particles
will have larger motions; likewise $\Delta$ is an average over
all particles, so that some particles are closer together.  Thus, in
practice, it is desirable that the MSFD is much less than
$\Delta^2$.

To quantify this last statement, we tracked simulated particles between frames with progressively larger
normalized MSFD, $\epsilon^2_n \equiv \langle dr^2_n \rangle / \Delta^2$, by increasing $n$, and tested how the
CG-algorithm performed when pushed past its original design parameters. For each $\epsilon^2_n$, we measured the
fraction $f$ of correctly tracked particles.  We checked in all cases that the vast majority of the tracking
errors are generated in the bulk of the system rather than at the boundaries.\footnote{We note the following
points: (1) The particle radius $R$ is not relevant to the tracking algorithm, and matters in the simulation only
as a `hard-particle constraint' and determines the density of the system via the area or volume fraction,
$\phi_{2D}$ or $\phi_{3D}$, respectively. (2) The tracking range $R_T$ used to test the CG-algorithm is the
largest possible beyond which combinatorics become excessive.}

\subsubsection{Quiescent system}
\label{subsubsec:quieslimit}

\begin{figure}
\includegraphics[width=0.45\textwidth,clip]{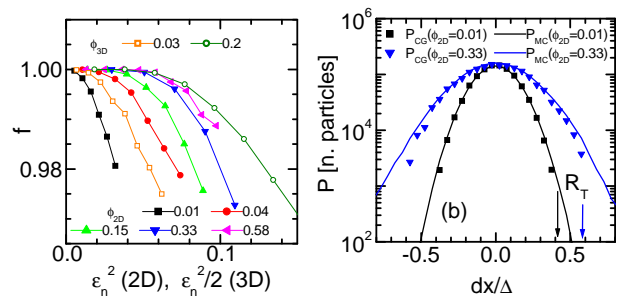}
\caption{(a) The fraction of correctly-tracked particles, $f$, versus the normalized true MSD $\epsilon_n^2$
between tracked frames for MC simulations of hard-disk fluids at various densities $\phi_{2D}$ and 3D hard-sphere
fluids at $\phi_{3D}=0.03$ and $\phi_{3D}=0.2$. Note the different $x$-axis for the 2D and the 3D case. (b)
Corresponding displacement distribution functions for two densities $\phi_{2D}$, taken for $f \simeq 0.98$. Solid
lines: the true PDF over $n$ MC steps. Symbols: $P_{CG}$ as obtained from classic CG tracking between frame $i$
and $i+n$, for $\phi=0.01$  ($\blacksquare$) and $\phi=0.33$ (\Blue{$\blacktriangledown$}).}\label{fig:8}
\end{figure}

In Fig.~\ref{fig:8}(a) we show $f$ as function of $\epsilon_n^2$ for the 2D system at different $\phi_{2D}$ and
$\phi_{3D}$. As expected, $f$ decreases with increasing $\epsilon_n$, but the performance in more concentrated
systems is considerably better than in the dilute case. To quantify this finding, we impose a criterion for
`successful tracking' of $f > 0.99$, and we find that the algorithm works up to $\epsilon_n = 0.15$ at the lowest
$\phi_{2D}$ studied, but this figure rises and essentially saturates at $\epsilon_n = 0.3$ at the highest
$\phi_{2D}$. This behavior reflects the difference in structure between a dilute and concentrated system. While
$\Delta$ is the average nearest neighbor spacing, in a dilute system, particles can approach much closer than this
(although still limited to be at least $2R$ apart). In this case, two closely spaced particles could potentially
swap positions and confuse the tracking algorithm. In contrast, for a concentrated system, $\Delta \approx 2R$,
and it is much harder for particles to swap positions. Thus in a concentrated system there are fewer
misidentifications for a given value of $\epsilon$ compared to a dilute system with the same $\epsilon$. For the
3D systems, the performance at large concentration is even better, i.e. for $\phi_{3D} =0.2$, the algorithm works
up to $\epsilon_n \simeq 0.4$ and is expected to saturate at $\epsilon_n \simeq 0.5$ for the largest $\phi_{3D}$.

An experimental diagnostic for correct tracking is the distribution of particle displacements from frame to frame,
$P(dx)$ \footnote{This is also known as the self-part of the van Hove correlation function.}. For correct tracking
it should  vanish smoothly within the tracking range. Figure \ref{fig:8}(b) compares the true distribution
function over $n$ frames, $P_{MC}$, with that resulting from the classic CG tracking between frames $i$ and $i+n$,
$P_{CG}$, for two densities and $n$ such that $f \simeq 0.98$. In both cases $P_{CG}$ follows the simulation data
for $dx<R_T$, beyond which is cut-off, and the discrepancies with $P_{MC}$, due to the misidentified particles
($1-f = 2$\%), appear in the large $dx/\Delta$ tails (clearer in the denser case). Note that the CG algorithm is
able to follow particles for larger tracking ranges in the case of larger area (or volume) fractions.

\subsubsection{2D system in shear flow}
\label{subsubsec:shearlimit}

\begin{figure}
\includegraphics[width=0.45\textwidth,clip]{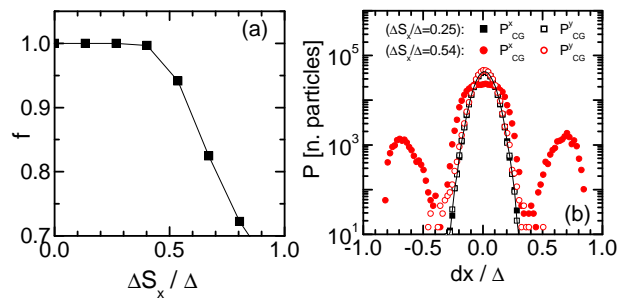}
\caption{Evaluation of classic CG tracking under shear: (a) $f$ versus the normalized maximum difference in
advected displacement between frames $\Delta S_x/\Delta=L_y \delta \gamma/\Delta$ for $\phi_{2D}=0.33$ and
$\epsilon^2=0.005$. (b) The distribution of particle displacements for $\Delta S_x/\Delta=0.25$ (squares) and
$\Delta S_x/\Delta=0.54$ (circles). Solid line: result without shear.}\label{fig:9}
\end{figure}

To test the performance of classic CG tracking
in the presence of non-uniform motion,
we superimpose affine shear and random displacements with a
MSFD of $(\epsilon \Delta)^2$ on a single 2D MC configuration
with $\phi_{2D}=0.33$. The strain increment between frames
is $\delta \gamma$, i.e.~the affine $x$-displacement over
one frame for particle $k$ is $\delta \gamma (y_k-\bar{y})$
(subtracting $\bar{y}$ guarantees zero net motion), again with periodic
boundary conditions. We analyze data only over accumulated
strains $< 20\%$, so that shear does not bring neighboring
particles in close proximity, avoiding `artificial' reduction
in performance (see Sec.~\ref{subsec:limitcorrection} and
Fig.~\ref{fig:15}).
The true origin for tracking errors
is the increase in the {\it difference} in advected displacements
between different parts of the `image', the maximum of which
is $\Delta S_x/\Delta=L_y \delta \gamma/2\Delta$ in units of
the average spacing, with $L_y$ the system size in the velocity
gradient direction. Figure~\ref{fig:9}(a) shows that $f$
rapidly decreases for $\Delta S_x/\Delta \gtrsim 0.4$.

From the
resulting tracks, we obtain the distribution of non-affine frame to
frame displacements after subtracting the
affine shear as evaluated from the classic CG
trajectories. The results for $x$ and $y$ displacements are shown
in Fig.~\ref{fig:9}(b). For $\Delta S_x/\Delta=0.25$ the
result is identical to that without shear, matching the superimposed
random motion. For $\Delta S_x/\Delta=0.54$, the distribution of
$y$-displacements appears very close to the correct distribution,
but the deformed central peak of $P^x$ and the presence of prominent `side bands' show that tracks have been evaluated incorrectly.

Thus, the classic CG algorithm can track particles between
consecutive frames for a maximum MSFD $\lesssim (0.3 \Delta)^2$ in
quiescent concentrated hard-sphere-like systems, but considerably less in
dilute systems.  These limits are similar to those discussed in the
original article by CG \cite{CrockerGrierJColIntSc96_tracking}.
Simply put, for larger displacements, the problem of uniquely
identifying particles becomes ill-posed, as the possibilities of
particles exchanging places become too significant.  No algorithm
can succeed in this case, and the only remedy is to acquire images
at a faster rate to resolve the intermediate steps.

Further, for
non-uniform flow the limit is set by a maximum {\it difference}
in advected motion of $\sim 0.4 \Delta$ over the full image. We
also found that cut-off and distortion effects in the distribution of particle displacements can
indeed be used as diagnostic of incorrect tracking, although
one should be cautious to interpret the absence of such features
as proof of $100\%$ performance.  The next subsection discusses
a simple modification of the classic CG algorithm to deal better with
particles in uniform or non-uniform flow.

\subsection{Iterated CG tracking algorithm}
\label{subsec:iterated}

\subsubsection{Description}
\label{subsubsec:iterated}

The classic CG algorithm was designed for cases where all particles move randomly (due to Brownian motion).
However, many interesting cases have particles moving in a flow with larger coherent structures, perhaps also with
Brownian motion superimposed, or even simply noise.  For example, the coherent motion could be due to Poiseuille
flow through a pipe, overall drift of the field of view, or an induced shear flow.  If the magnitude of this
motion is small, the classic CG algorithm still has some ability to track particles. Following the logic above,
tracking should work reasonably well if the distance most particles move between frames is moderately less than
the interparticle spacing $\Delta$, whatever the origin of this motion may be.

In cases where the motion is simple and small compared to $\Delta$, the classic CG algorithm can be iterated to
produce better results. This ``Iterated Tracking'' method is as follows.  (i) First identify the particle
positions at each time, as per Sec.~\ref{sec:Locating}. (ii) Track the particles using the classic CG algorithm.
(iii) Determine the coherent motion from the successfully tracked particles.  (iv) Remove the coherent motion from
the original particle positions.  (v) Repeat steps ii-iv until most particles are successfully tracked, and the
residual coherent motion detected in step iii is reduced to an acceptable level.  (vi) Add back in all of the
coherent motion that has been previously subtracted in all iterations of step iv.

As long as the motion of most particles is less than the tracking
limit $R_T$ and thus less than the interparticle spacing
$\Delta$, at least a few particles will be successfully tracked
in the first iteration.  The coherent motion of these few
particles is then used to `bootstrap' the classic CG algorithm,
and in the subsequent iterations, more particles are correctly
tracked.  These then refine the coherent motion and thus the
iterated tracking method eventually is able to converge on the
correct trajectories for all the particles.  In practice, this
usually only takes 3-4 iterations to produce good results.  The key to iterated
tracking is that the first tracking step must {\it correctly}
track enough particles to start the process.  We use our
simulated data to study the breakdown of iterated tracking in
a test case.

\subsubsection{Uniformly moving system}
\label{subsubsec:unilimit}

\begin{figure}
\includegraphics[width=0.45\textwidth,clip]{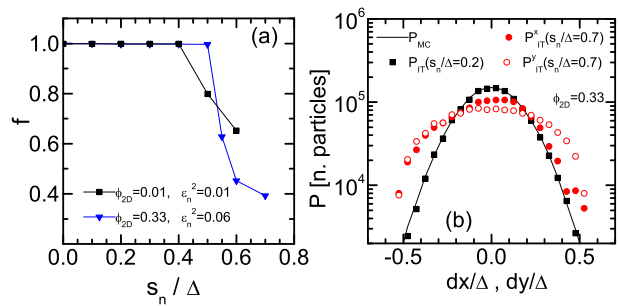}
\caption{Iterated tracking results.  (a) $f$ versus the normalized shift $s_n/\Delta$ between frames for
$\phi=0.01$, $\epsilon_n^2=0.01$ and $\phi=0.33$, $\epsilon_n^2=0.06$. (b) PDF's in the co-moving frame for
$\phi_{2D}=0.33$ and two `drift' velocities: Line: the true PDF over $n$ MC frames. ($\blacksquare$): $P_{CG}$ for
$s_n/\Delta=0.2$, (\Red{$\bullet$}): distribution function of $x$-displacements $P^x_{CG}$, in the co-moving frame
for $s_n/\Delta=0.7$, (\Red{$\circ$}) same for $y$-displacements.}\label{fig:10}
\end{figure}

We first superimpose a uniform $x$-displacement $s_n$ over $n$ MC steps on top of the MC dynamics. Periodic
boundary conditions keep particles within the analysis window. We then perform iterated tracking on these particle
positions.  Figure \ref{fig:10}(a) shows $f$ versus $s_n$ for two densities ($f(s_n=0)>0.995$ in both cases). For
small $s_n$, $f \simeq f(s_n=0) \approx 1$, but for $s_n/\Delta \gtrsim 0.5$ very few correct tracks are found.
Note that iterated tracking still provides a result, but mostly consisting of incorrect tracks, and yields
incorrect motion; in this case, the first tracking step has failed and subsequent tracking steps are unable to
improve the results. In Fig. \ref{fig:10}(b) we show the distribution of particle displacements in the {\it
co-moving frame} for two values of $s_n/\Delta$. By construction, the true $P_{MC}$ is identical to that in the
quiescent system, while $P_{IT}$ is the displacement distribution in the co-moving frame, i.e.~the frame where the
average displacement of the `IT-tracks' (obtained via the iterated tracking algorithm) vanishes. The breakdown of
iterated tracking for $s_n/\Delta=0.7$ in the $\phi=0.33$ system is brought out by the sharp cut-off of
$P(dx>R_T)$, and, more prominently, by the asymmetry in the distribution of displacements $P^x_{IT}$ along the
direction of motion.

The results are similar for non-uniform flow.  If the maximum motion in any area of the data exceeds $\sim 0.4
\Delta$, that region will be poorly tracked.  For example, with the shearing data discussed in
Sec.~\ref{subsubsec:shearlimit}, at each iteration step the measured shear strain can be removed, but this method
will still have difficulties when $\Delta S_x \gtrsim 0.4$.  With care, trajectories found in regions with motion
less than $0.4 \Delta$ may be used to extrapolate the motion to the more mobile regions. Note that the classic CG
algorithm was not intended for tracking particles in flow, so that the success of iterated tracking in non-trivial cases reveals the strength of that algorithm.

\subsection{Correlated image tracking}
\label{subsec:cit}

\subsubsection{Description}
\label{subsubsec:citdescription}

Due to the limitations shown above, 2D or 3D images with
large drift or non-uniform motion require a modified
analysis for correct tracking. Similar shortcomings of
standard tracking for granular flows have been discussed in
\cite{XuReeveseRSciInstr04_trackingerrors} and analysis of such
data will also benefit from the correction method we describe
in this section. The basic ingredients of ``Correlated Image
Tracking'' are:
(i) We first obtain the particle coordinates as explained in
Sec.~\ref{sec:Locating}. (ii) We then obtain independent information
on the advected motion and its spatial and time dependence via
PIV-type correlation analysis of the raw images. (iii) Next,
this (time and position dependent) advected motion is subtracted
from the bare particle coordinates. This yields the particle
coordinates in a `locally co-moving' (`CM') frame. (iv) In the CM
frame, the particles can be tracked as in a quiescent system. The
tracking efficiency is essentially limited by the value of the
MSFD or non-affine motion in the CM frame. (v) After tracking,
the advected motion is added back to the particle coordinates to
obtain the trajectories in the laboratory frame.

\begin{figure}
\includegraphics[width=0.45\textwidth,clip]{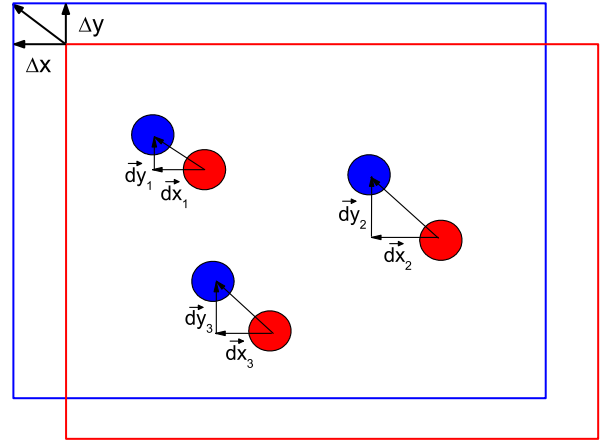}
\caption{Illustration of the shifting and correlation procedure. Each particle in the section of the image under
consideration is displaced by $(dx,dy)$ between frame 1 and frame 2. The entire image section is shifted over
$(\Delta x, \Delta y)$ to maximize the correlation between the two successive sections.}\label{fig:11}
\end{figure}

To identify frame to frame advective motion, we use standard PIV-type image correlation methods ~\cite{PIV}.
Consider a region of size $n \times m$ pixels of two consecutive 2D images $i-1$ and $i$. Let $I_{i-1}(x,y)$ and
$I_{i}(x,y)$ be the intensity patterns as function of position $(x,y)$ of these (sections of) images. The
covariance is defined as:
\begin{eqnarray}
cov[I_{i-1}(x,y),I_{i}(x,y)]=~~~~~~~~~~~~~~~~~~~~~~~~~~ \nonumber\\
\displaystyle{\frac{1}{[(n \times m) - 1]} \sum_{p=1}^{n}\sum_{q=1}^{m} \left[ I_{i-1}^{pq} - \langle I_{i-1}
\rangle \right] \left[ I_{i}^{pq} - \langle I_{i} \rangle \right]},
\end{eqnarray}
where $I_{i-1}^{pq}$ and $I_{i}^{pq}$ are the intensities of the pixel corresponding to position $(x_p,y_q)$ in
each image and $\langle I_{i-1} \rangle$,$\langle I_{i} \rangle$ are the respective average intensities defined as
$\langle I \rangle = \displaystyle{\frac{1}{[n \times m]} \sum_{p=1}^{n}\sum_{q=1}^{m} I^{pq}}$. Analogously,
the variance of a single image $I$ is:
\begin{equation}
var[I(x,y)] = \displaystyle{\frac{1}{[(n \times m)-1]} \sum_{i=0}^{n-1}\sum_{j=0}^{m-1} (I^{ij} - \langle I
\rangle)^2}.
\end{equation}
The correlation coefficient $c[I_{i-1}(x,y),I_{i}(x,y)]$ of the two images is defined as:
\begin{equation}
c[I_{i-1}(x,y),I_{i}(x,y)] =
\displaystyle\frac{cov[I_{i-1}(x,y)I_{i}(x,y)]}{\sqrt{var[I_{i-1}(x,y)]var[I_{i}(x,y)]}}.
\end{equation}
The motion is obtained by shifting image $i$ by a certain number
of pixels $(\delta x,\delta y)$ and computing the correlation
coefficient $c[I_{i-1}(x,y),I_{i}(x - \delta x,y -\delta y)]$. This
is repeated for shifts within a desired range until $c$ is
maximized for $\delta x=\Delta x$, $\delta y=\Delta y$. Repeating
the procedure over subsequent frames yields the displacement as
function of time $(\Delta x,\Delta y)(t_i)$ in the region of the
image series under consideration, Fig.~\ref{fig:11}. For strongly time-dependent
flows, a scan over all possible $\delta x$, $\delta y$ in this
region is required, but for smooth flows we have implemented a
more efficient method in which image $i$ is scanned in a narrow
range centered around the shift $(\Delta x, \Delta y)(t_{i-1})$
found from the previous images.  A key point to recognize is that
these shifts are quantized by the size of a pixel, so while the
motion obtained from image correlation is a good starting point
for the tracking, the subsequent tracking is necessary to achieve
subpixel resolution of particle motion.

In most applications one can identify, at least within the microscope field of view (see
Sec.~\ref{subsec:cfrheoscope}), a principal axis along which the flow takes place. The entire image can then be
rotated such that the advective motion occurs in only one direction, which we denote by $x$. To obtain the
advection profile over the entire image, the correlation method is applied in different ways depending on the
uniformity of the motion and image dimensionality, as we describe now.

\begin{figure}
\includegraphics[width=0.45\textwidth,clip]{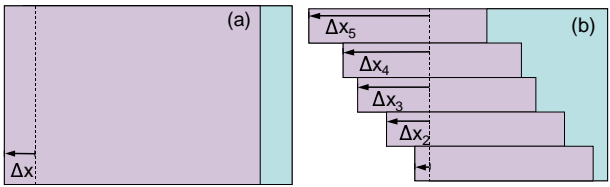}
\caption{Examples of 2D image correlation procedures. (a) A uniform shift $\Delta x$ across the entire field of
view maximizes the correlation. (b) The advected motion is a function of $y$; the image is then decomposed in bins
centered at $y_q$, each of which is shifted by $\Delta x_(y_q)$ to obtain maximum correlation.}\label{fig:12}
\end{figure}

For 2D images, when the motion is spatially uniform in the $xy$ plane, the above procedure is applied to the
entire image, Fig.~\ref{fig:12}(a), resulting in $\Delta { \bf r}({\bf r},t_i)=\Delta x({\bf
r},t_i)=\Delta x(t_i)$. The advective motion can also depend on the position $y$ transverse to the flow, see
Fig.~\ref{fig:12}(b). The image is then decomposed into strips, which are shifted and correlated
separately, yielding an advection profile $\Delta x(y_q)$ discretized at the centers $y_q$ of the strips.

\begin{figure}
\includegraphics[width=0.45\textwidth,clip]{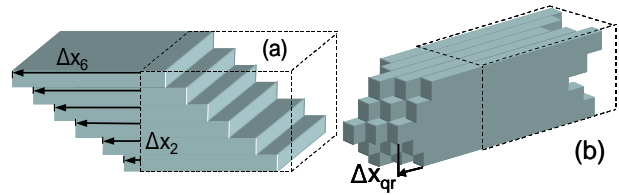}
\caption{Examples of 3D image correlation procedures. (a) The motion is a function of $z$ only. The 3D image is
decomposed in slices centered at $z_r$, each of which is shifted by $\Delta x(z_r)$ to obtain maximum correlation.
(b) The advected motion is a function of $y$ and $z$. Decomposition into $y$ and $z$ bins yields the advection
profile $\Delta x(y_q,z_r)$.}\label{fig:13}
\end{figure}

For 3D images, the basic manifestation of nonuniform flow is simple shear, Fig.~\ref{fig:13}(a), where
the average motion is a function of $z$ only. The sheared volume is then decomposed in $xy$ slices at different
$z$ and the correlation procedure is performed on each 2D slice separately. A more complex flow is shown in
Fig.~\ref{fig:13}(b), where shear occurs both in the $y$ and $z$ direction, as for instance in 3D
channel flows. Here the 3D images are first decomposed in $xy$ slices at different $z$ and then each slice is
further decomposed in $y$-bins for which the motion is analyzed.

In the most general form ${\mathbf{\Delta r}}$ is both position and time dependent, ${\mathbf{\Delta r=\Delta r(
r}},t_i)$, and includes shifts in all three directions $\Delta x({\mathbf{r}},t_i), \Delta y ({\mathbf{
r}},t_i),\Delta z({\mathbf{r}},t_i)$. For example, in experiments where a point-like force source is applied in
the medium, e.g. by dragging a magnetic or tracer bead through a colloidal suspension~\cite{habdas}, the direction
and the magnitude of the `advected' motion depend on $x$, $y$ and $z$. In such a case the imaged area (volume)
must be decomposed into squares (cubes), and a full PIV analysis must be carried out to characterize the motion
$\Delta {\bf r}(x_p,y_q,z_r,t_i)$ in each element $p,q,r$. Another example is sedimentation, where $\Delta x=0$,
$\Delta y=0$ but $\Delta z$ depends on $z$. Extensions to our algorithm dealing with such cases are possible but
we have not implemented this. For our experiments (simple shear or channel flow) it suffices to consider shifts in
one direction, which are independent of the coordinate in that direction.

Once the advected motion is measured, the discrete displacement
profile $\Delta {\bf r}(x_p,y_q,z_r,t_i)$ is then interpolated to
give the continuous profile $\Delta {\bf r}(x,y,z,t_i)$.  This
is more appropriate to subtract from the particle coordinates, which are themselves continuously distributed.
Using the continuous profile, the transformation of the position
$\mathbf{r}_k(t_i)=[x_k,y_k,z_k](t_i)$ of particle $k$ in the
laboratory frame to its position ${\mathbf{\bar{r}}}_k(t_i)$
in the CM frame of reference is:
\begin{equation}
{\mathbf{\bar{r}}}_k(t_i) = {\mathbf{r}}_k(t_i) - \sum _{j=1}^{i}
{\mathbf{\Delta r( r_k(t_i)}},t_j) ,
\label{eq:general-advection-removal}
\end{equation}
where ${\mathbf{\Delta r}}( {\mathbf{r}}_k(t_i),t_j)$ is the {\it past} motion between frame $j$ and $j-1$, at the
{\it current} location ${\bf r}_k(t_i)$ of the particle.  This reduces to $\Delta {\bf r}({\bf r},t_j)=\Delta {\bf
x}(y,z,t_j)$ for our experiments.

In the CM frame, the average particle motion (nearly) vanishes. The use of the classic CG algorithm at this point
therefore allows particle tracking limited only by the MSFD in the CM frame. Occasionally, the CM tracks show some
residual motion, in which case a modest improvement may be obtained by using Iterated Tracking
(Sec.~\ref{subsec:iterated}).

Once tracking is completed, the advected motion $\Delta {\bf r}$
is added back by inverting Eq.~\ref{eq:general-advection-removal}.
This then provides the trajectories of the particles in the
laboratory frame of reference.

\subsubsection{Limitations of Correlated Image Tracking}
\label{subsec:limitcorrection}

\begin{figure}
\includegraphics[width=0.45\textwidth,clip]{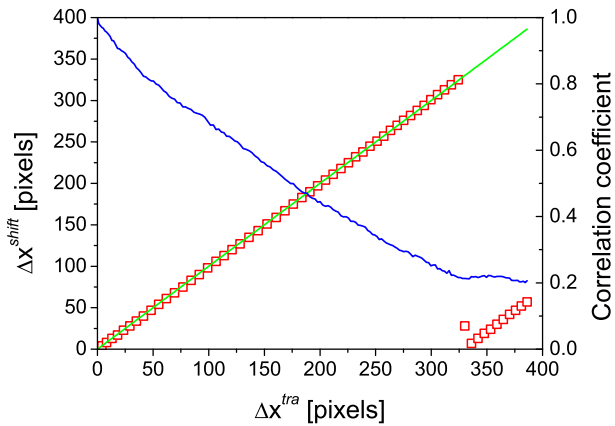}
\caption{Test of the shift-correlation procedure on images of plug-flow. The shift in the flow direction $\Delta
x^{shift}$ (\Red{$\blacksquare$}), obtained from the correlation method, is plotted against the exact accumulated
displacement obtained from tracking $\Delta x^{tra}$ (\Green{--}), see the text. Image size: $331~\times~580$
pixels. Right axis: correlation coefficient (blue line) versus $\Delta x^{tra}$.}\label{fig:14}
\end{figure}

A possible limitation to correlated image tracking is a failure of the PIV-type correlation method. This could
arise when the frame to frame shifts are a significant fraction of the actual image or image-bin size. However,
this method can work successfully for quite large shifts amounting to nearly the image size.  To show this, we
analyzed an experimental image series taken from plug flow.  In this image series, the motion was slow enough that
classic CG tracking worked; we use the results of this tracking as the ``true'' motion. Furthermore, there was
little relative particle motion (nearly zero MSFD).  We then took pairs of images from this image series,
separated by $n$ frames, to model an effectively much faster flow rate.  From the tracking, we know the shift
$\Delta x^{tra}$ that should best align these two images.  For each pair of images, we calculate the shift $\Delta
x^{shift}$ from image correlation, and compare that apparent shift to the true shift $\Delta x^{tra}$, in
Fig.~\ref{fig:14} (red symbols).  The correlation coefficient (thin blue line) decays roughly linearly with
$\Delta x^{tra}$, in line with the reduction of the correlated portion of the images, and plateaus at a value
$\sim 0.2$, corresponding to the coefficient for two entirely different images of the same system. The latter
value is specific to our high density system, and may vary for different systems.

When comparing the image at time $j$ and the shifted image at time $j+1$, it is important to note that in each
case the full image is compared.  Thus the shifted image at $j+1$ has some pixels wrapped around from one edge of
the image to the other, which has no physical meaning.  An alternate idea would be to crop the two images, so that
any pixels shifted outside the boundary are removed. Thus, when considering very large shift factors approaching
the width of the image, only two narrow strips of the two images would be compared to determine the correlation
coefficient. We find that this method is generally less successful, despite its intuitive appeal.  When the
required shifts are large, generally using the full image is more likely to find the correct shift value. Comparing the large regions of the two images that could be potentially cropped, these will be uncorrelated, and thus in general the correlation coefficient is dominated by the small regions that correctly overlap. This then results in the
results of Fig.~\ref{fig:14} where the correct shift value is found even for $\Delta x^{tra}$ almost
as much as the full width of the image.

\begin{figure}
\includegraphics[width=0.45\textwidth,clip]{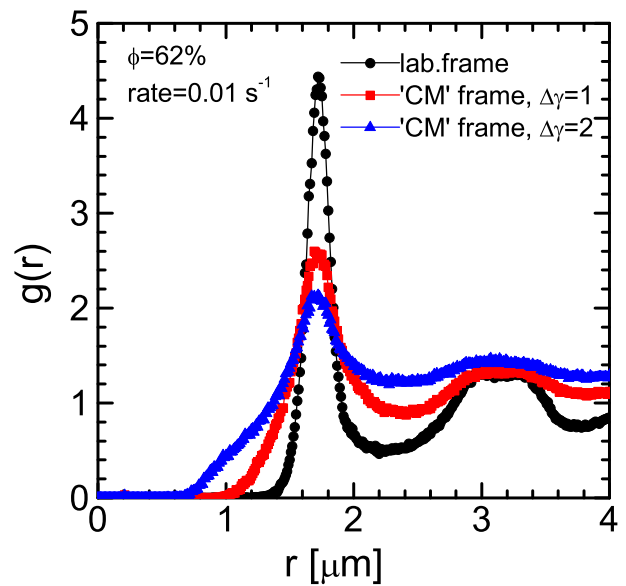}
\caption{Short distance behavior of the pair correlation function $g(r)$ from 3D coordinates of a sheared
colloidal glass ($\phi=0.62$, $R=850$~nm), in the laboratory frame ($\bullet$), in the locally co-moving (`CM')
frame after subtracting advected motion corresponding to an accumulated strain of $\Delta \gamma=100 \%$
(\Red{$\blacksquare$}), and $\Delta \gamma=200\%$ (\Blue{$\blacktriangle$}). The local shear rate is
$\dot\gamma=0.01$~s$^{-1}$, strain accumulation during acquisition of a single 3D stack is $\sim 1.5\%$, while
between frames it is $4 \%$. The data show that particles may come in close proximity after removal of large
advected motion.} \label{fig:15}
\end{figure}

A stronger limitation to the correlation procedure is an excessive
amount of {\it relative} particle motion between frames. This
obviously limits the correlation compared to that of plug flow
described above, but, more directly, large relative displacements
cause failure of classic CG tracking in the CM frame, as discussed
in Sec.~\ref{subsubsec:quieslimit}. In practice we found that
a maximum MSFD in the co-moving frame (the non-affine motion) of
$\sim (0.3\Delta)^2$ is the limiting value for the method to work.

Non-affine motion during flow has an additional effect which may limit the tracking performance when subtracting
shear advection accumulated over many frames. Suppose two particles start off as nearest neighbors, at a distance
$\geq 2R$, but at `streamlines' with different velocities. After one frame, when the advected motion is
subtracted, they remain at that distance if their motion was fully affine. In contrast, {\it with} non-affine
motion, their separation after advection removal may become $<2R$, that is, they would apparently be in contact.
Over single frames this effect is limited, i.e.~particle separations still considerably exceed their non-affine
displacements between frames and tracking is not affected. However, after multiple frames, the non-affine
displacements can accumulate and the particle separation in the CM frame may have been reduced to a value
comparable to the non-affine frame to frame displacements. This gives rise to tracking errors based on the same
arguments as in Sec.~\ref{subsubsec:quieslimit}. The artificial reduction in particle separation in the CM frame
is visible in the pair correlation function $g(r)$, Fig.~\ref{fig:15}, where we see that some particles apparently
overlap after the removal of large advective motion.

This problem is remedied by {\it piecewise tracking} of particles
in intervals over which the accumulated relative motion is small.
Each interval the particles are assigned their identification
tag, and the full trajectory is obtained by matching the
tags in a one-frame overlap of the intervals. For the case in
Fig.~\ref{fig:15}, particle separations in the co-moving
frame are $\geq R$ for accumulated strains $\Delta \gamma \leq
100\%$, so that tracking is only affected for $\Delta \gamma
> 100\%$, but in general a different limit may apply since
non-affine motion does not necessarily scale with accumulated
strain \cite{BesselingPRL2007} and may depend on $\phi$.

Summarizing, the main advantage of correlated image tracking is that particles are tracked in a {\it locally
co-moving frame of reference} where limitations to tracking are the same as in a quiescent system, as described in
Sec.~\ref{subsubsec:quieslimit}. In other words, our method permits the tracking of particles in flowing colloids
to the same level of accuracy as that in non-flowing systems.

\section{Applications}
\label{sec:Examp}

In this section, we give example results from particle tracking in the two flow geometries already introduced in
Sec.~\ref{sec:Instrument}, viz., simple shear and capillary flow. We also describe the application of the confocal
rheoscope to perform simultaneous rheology and velocity profiling of soft materials.

\subsection{2D tracking of channel flows}
\label{subsec:2dchannel}

We start with a 2D example, the characterization of pressure-driven channel flow of colloidal pastes. We used a
$\phi \approx 0.63$ suspension of fluorescent PMMA spheres (radius $R = 1.3 \pm 0.1$~$\mu$m, from microscopy),
suspended in a mixture of CHB and mixed decalin for refractive index and buoyancy matching. A pressure difference,
$\Delta P$, was applied to drive the suspension into the square channels (side $a = 50$~$\mu$m, smooth inner
walls). The flow across the full channel width was imaged in 2D at 107 frames per second (image size $44 \times 58
\mu{\rm m}^2$, $256 \times 320$ pixels) at a depth of 17 $\mu$m from the lower surface. The images were collected
at a distance corresponding to $\sim 2000$ particle diameters from the channel inlet where entry effects have died
out and the flow has negligible $x$-dependence on a scale compared to image size. The particles are located in 2D
with accuracy $\sim 50$~nm ~\cite{CrockerGrierJColIntSc96_tracking,isa1}. According to
Eq.~\ref{eq:distortionlimit}, distortion of the particle image becomes significant only for $V > V^{max}_{2D}
\simeq 600$~$\mu$m/s, which exceeds the maximum velocities at which we are able to track the particle. Assuming
uniform motion, the results in Sec.~\ref{subsubsec:unilimit} suggest that iterated tracking fails for flow
velocities $V>V^{max}\simeq R \times f_{\rm scan} \simeq 140$~$\mu$m/s, but in practice the limit is smaller due
to the presence of large velocity gradients.

\begin{figure}
\includegraphics[width=0.45\textwidth,clip]{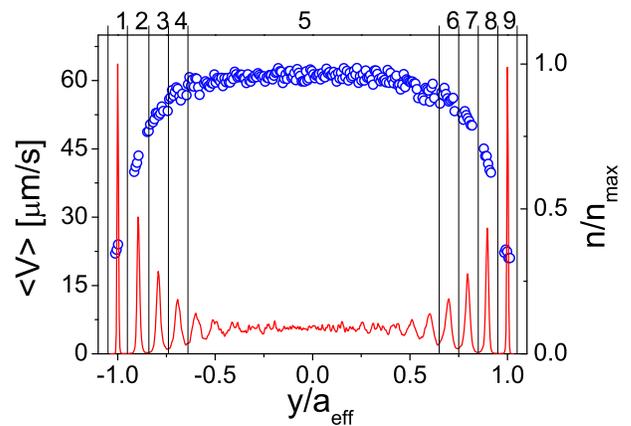}
\caption{Velocity profile (\Blue{$\bullet$},left axis) and histogram of particle positions (\Red{--},right axis)
as a function of the transverse coordinate $y$ normalized by effective channel size $a_{\rm eff} = a-R$. The
experiment refers to the flow of a 63.5\% suspension in a $a=50 \mu$m wide smooth, square channel. The particles
are arranged in well defined layers close to the walls and the shear decays towards the channel center where the
suspension flows as a plug. The vertical lines highlight the $y$-bins and the numbers match those in
Fig.~\ref{fig:17}.}\label{fig:16}
\end{figure}

For the experiments considered here, the flow profiles consist of a central region of uniform velocity $V \simeq
V_c$ and lateral zones adjacent to the channel walls where the shear is localized,
Fig.~\ref{fig:16}. We have discussed the detailed physics elsewhere~\cite{isa2}. To comment briefly, we note that such a profile in itself,
could be consistent with predictions from the rheology of yield stress fluids~\cite{You05}. However, in contrast
to yield stress fluid predictions, the velocity profiles scale with flow rate and the width of the shear zones is
independent of flow rate ~\cite{isa2}. By examining the microscopic dynamics of the particles we observe that they
are dominated by interparticle collisions and contacts with similarities to dense granular flows. The similarities
extend to the shape and scaling of the velocity profiles, which allowed to interpret the data using a stress
fluctuation model initially conceived for dry grains~\cite{isa2}.

\begin{figure}
\includegraphics[width=0.45\textwidth,clip]{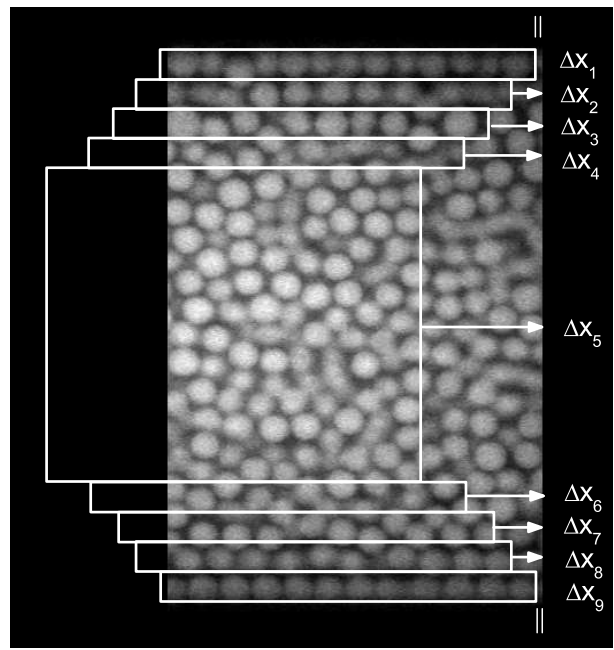}
\caption{Illustration of the non-uniform shifts during 2D channel flows. The resulting displacements $\Delta x_i$
are subtracted from the particle coordinates. Image size: $34~\mu$m $\times~50~\mu$m.}\label{fig:17}
\end{figure}

\begin{figure}
\includegraphics[width=0.4\textwidth,clip]{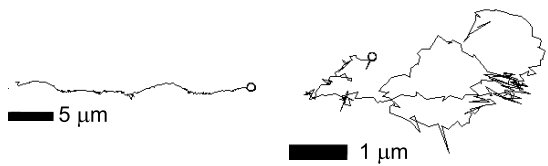}
\caption{Example of particle tracks in the laboratory (left) and in the co-moving frame (right). Circles indicate
the start of the trajectory. The particle was situated in a shear zone (local shear rate $5 $~s$^{-1}$) of a
suspension flowing into a square channel with smooth inner walls. The length of the track is $\sim
2.3$~s.}\label{fig:18}
\end{figure}

In these experiments, the advected motion $\Delta x(y)$ is analyzed
using correlated image tracking with a scheme similar to that in
Fig.~\ref{fig:12}(b). We divide the image in horizontal
bins with sufficiently uniform displacements and obtain $\Delta
x(y_q)$ from correlation in each of these. For channels with smooth
walls, the particles near the wall are arranged in well defined
layers, Fig.~\ref{fig:16}. Combined with the fact
that velocity gradients are largest near the edge, this motivates
the choice of one-particle-wide bins near the walls and a larger
bin in the center. An example of the displacement profile $\Delta
x(y_q,t_i)$ is shown in Fig.~\ref{fig:17}
and, combined with Eq.~\ref{eq:general-advection-removal}, can be
used to track the particles in the CM frame. Fig~\ref{fig:18}
shows a particle trajectory inside a shear zone both in the
laboratory (left) and in the CM frame (right).

\begin{figure}
\includegraphics[width=0.45\textwidth,clip]{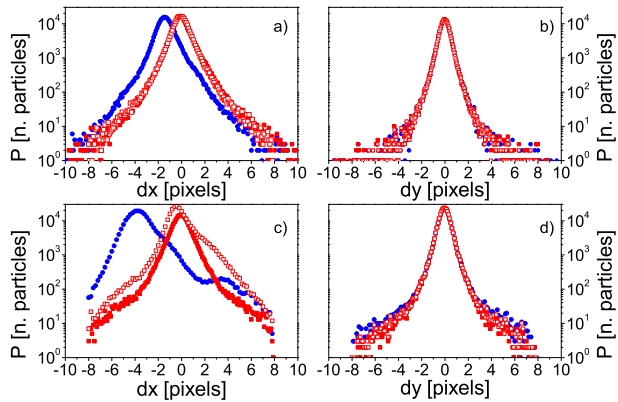}
\caption{Comparison of displacement distributions along $x$ (a),(c) and $y$ (b),(d) for 2D channels flows. Classic
CG tracking in the laboratory frame (\Blue{$\bullet$}), Iterated CG tracking (\Red{$\square$}), Correlated Image
Tracking (\Red{$\blacksquare$}). (a),(b): slow flow ($\tilde{V}_c=-1.5$~pixel/frame); classic CG tracking is
sufficient. (c),(d): fast flow ($\tilde{V}_c=-12$~pixel/frame). Here the classic CG tracking is inapplicable in
the lab frame of reference (\Blue{$\bullet$}), but correlated image tracking utilizing CG tracking in the
co-moving frame of reference is successful.} \label{fig:19}
\end{figure}

In Figs.~\ref{fig:19}(a),(b) we first show the displacement distribution functions for a small velocity $V_c \simeq
28$~$\mu$m/s ($1.5$ pixel/frame), for which particles can be tracked directly using classic CG tracking. These
vanish smoothly within the tracking range and are consistent with the results (in the co-moving frame) from
iterated tracking (open squares) and with the results from tracking in the CM frame, i.e.~after removing
non-uniform motion (filled squares).

For faster flow, $V_c \simeq 223$~$\mu$m/s ($\sim 12$
pixels/frame), correlated image tracking is required.  In
Figs.~\ref{fig:19}(c) we show $P(dx)$ from classic
CG tracking (filled circles). The strong asymmetry and sharp cutoff show
the inapplicability of this method. Results from iterative tracking,
i.e.~after removing uniform motion obtained from a preceding direct
tracking step, (open squares) still reveal asymmetry and are not
reliable. Instead, the results from correlated image tracking (full
squares) are symmetric and show virtually no cut-off effects. The
$y$-dependent motion after restoring these tracks to the laboratory
frame are consistent with the advection profile $\Delta x(y)$
from correlation, confirming the success of the method. We note
that for the different methods, the distributions $P(dy)$, shown in
Figs.~\ref{fig:19}(d), do not show any difference
despite the variations observed in $P(dx)$. This again illustrates
that care must taken in interpreting displacement distribution functions from particle tracking
as described in Sec.~\ref{subsec:evaluation}.

Finally, we also tested the method on $\simeq 30 \%$ volume fraction suspensions in quasi-2D channels yielding the
expected parabolic flow profiles~\cite{isa1,frank03}. With correlated image tracking we have been able to
successfully track particles in flows with velocity as high as $250 \mu$m/s corresponding to almost twice the
limit $V^{max}$ mentioned above.

\subsection{3D particle tracking in simple shear flow}
\label{subsec:3Dtrack}

Next, we move to full 3D imaging and consider a colloidal glass in steady shear flow, measured in the rheometer or
the shear cell with rough, coated surfaces. The apparatus has already been described in Sec.~\ref{sec:Instrument}.
The colloids (radius $R=850$~nm) are suspended in a charge screened CHB-decalin mixture for refractive index and
density matching ($\eta=2.6$~mPa$\cdot$s). The volume fraction, measured from the average Voronoi volume
determined from particle coordinates, is $\phi=0.62$.  Each 3D image consists of $76$ slices ($256 \times 256$
pixels each, 2D frame rate $f=45$~s$^{-1}$), imaged over a height $z_{max}-z_0=15$~$\mu$m with $z_0$ either
$10$~$\mu$m or $15$~$\mu$m ($z=0$ at the cover slide) and was acquired in $\sim 1.7$~s. The voxel size is $0.11
\times 0.11 \times 0.20$~$\mu{\rm m}^3$. The local shear rate $\dot\gamma$ which we measure (see below) may exceed
the overall applied rate $\dot \gamma_a$. This is due to global shear localization, which we observed directly in
velocity profiles $v(z)$ measured from image series on a coarser $z$-scale, which we will describe in detail
elsewhere \cite{BesselingTBP}.

\begin{figure}
\includegraphics[width=0.45\textwidth,clip]{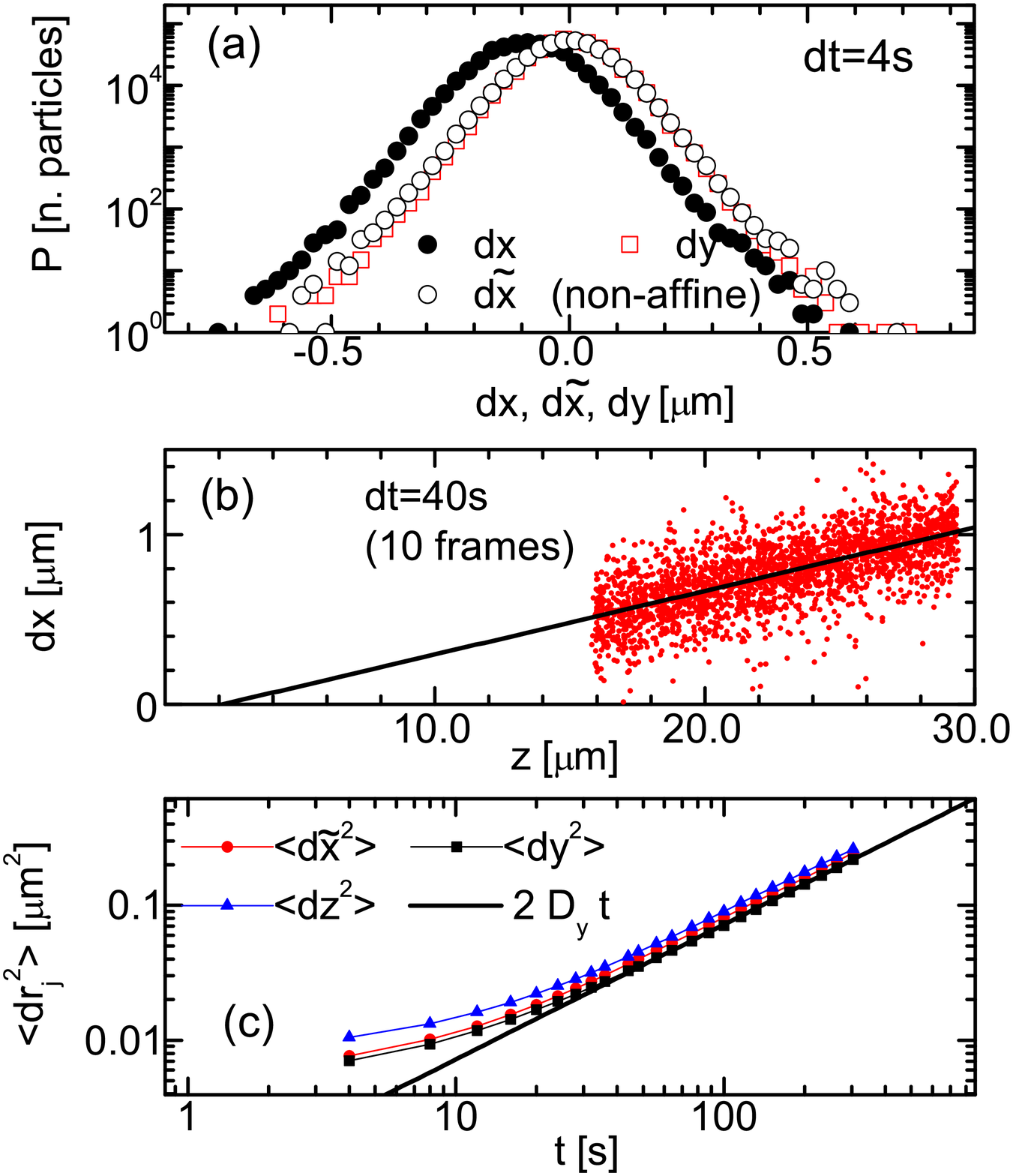}
\caption{Direct tracking of a sheared glass at $\dot\gamma=9.3 \times 10^{-4}$~s$^{-1}$ (a) Histograms of frame to
frame displacements. (b) Displacements $dx$ versus $z$ for all particles over $10$ frames ($dt=40$~s). The line is
a linear fit, the slope of which gives the local shear rate $\dot\gamma=9.3 \times 10^{-4}$~s$^{-1}$. (c) Mean
squared displacement $\langle dy^2(t) \rangle$ in the vorticity direction, $\langle dz^2(t) \rangle$ in the
gradient direction, and the non-affine MSD $\langle d\tilde{x}^2(t) \rangle$ in the velocity direction. Line:
$\langle dy^2(t) \rangle=2D_y t$ with $D_y=3.6 \times 10^{-4}$~$\mu{\rm m}^2$/s.} \label{fig:20}
\end{figure}

The intrinsic accuracy for locating particles (with the refinement in~\cite{JenkinsEgelhaafAdvColIntSci08}),
obtained from the MSD in a random close packed system (no flow) is $\pm 30$~nm in $x$,$y$ and $\pm 70$~nm in the
$z$-direction. Under flow, using Eq.~\ref{eq:distortionlimit} in Sec.~\ref{subsec:loclimit}, the velocity for
which distortion of the particle image sets in is $V^{max}_{3D} \sim 1$~$\mu$m/s, exceeding our largest measured
velocities $V^{max} \sim \dot\gamma z_{max}$. The error due to short time thermal displacement is $\pm 30$~nm, not
exceeding those mentioned above.

In Fig.~\ref{fig:20} we show results
for slow shear, where classic CG tracking is sufficient.
Figure~\ref{fig:20}(a) shows the distribution of
frame to frame displacements along $x$ (the velocity direction)
and $y$ (the vorticity direction). The former is shifted and slightly
broader compared to $P(dy)$ due to the $z$-dependence of $dx$
and the zero-velocity plane being outside the image, which is
illustrated by the displacement profile $\langle dx(z) \rangle$ in
Fig~\ref{fig:20}(b). The profile is linear on average on
this $z$-scale, the slope gives the local shear rate $\dot\gamma=9.3
\times 10^{-4}$~s$^{-1}$, and it extrapolates to zero within experimental uncertainties
at the cover-slide ($z=0$), confirming that the coating provides
a stick boundary condition.

\begin{figure}
\includegraphics[width=0.45\textwidth,clip]{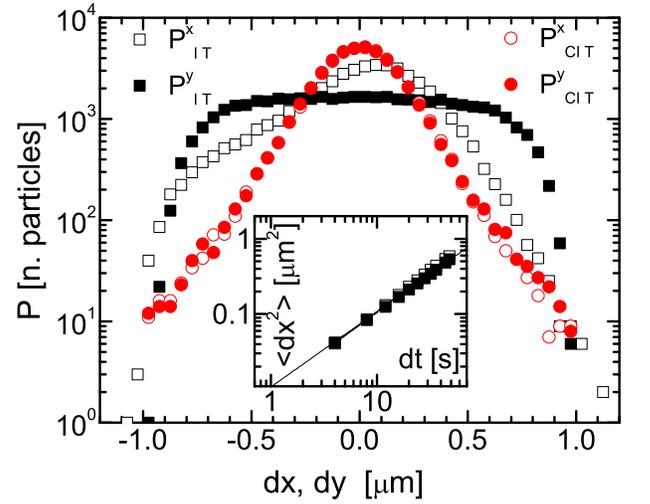}
\caption{Histograms of frame to frame displacements, in a comoving frame with $\langle dx \rangle=0$, obtained
from iterated CG tracking (${\rm P_{IT}}$, squares) and from correlated image tracking (${\rm P_{CIT}}$, circles)
for 2D imaging in the velocity vorticity plane of a sheared glass ($V=1.6$~$\mu$m/frame, $\dot
\gamma=0.019$~s$^{-1}$). Inset: the MSD $\langle dy^2(t) \rangle$ ($\blacksquare$), and $\langle dx^2(t) \rangle$
in the co-moving frame ($\square$), from tracking in the CM frame. Line: $\langle dy^2(t) \rangle=2Dt$ with $D=5.4
\times 10^{-3}$~$\mu{\rm m}^2$/s.} \label{fig:21}
\end{figure}

Locally, the shear induces plastic
breaking of the particle cages, causing diffusive behavior at long
times, as shown by the MSDs in Fig~\ref{fig:20}(c)
(see also \cite{BesselingPRL2007}). This diffusion contrasts a
{\it quiescent} colloidal glass where the long time particle
dynamics remains caged. Figure~\ref{fig:20}(c)
includes the MSDs in the three directions $x$, $y$ and $z$,
where for the $x$ direction we use $\langle d\tilde{x}^2(t)
\rangle$, with \begin{equation} d \tilde{x}(t)=x(t)-x(0)-\dot
\gamma \int_0^t z(t') dt', \label{eq:nonaffinedx} \end{equation}
which represents only the non-affine displacement. Note that
with this definition, the usual effect of Taylor dispersion is
suppressed, see e.g. \cite{YamamotoOnukiPRE98_SCLiq_rheodif}. As
observed, the MSDs are nearly isotropic, i.e.~the (non-affine)
structural relaxation due to cage breaking is nearly the same
for all directions \cite{BesselingPRL2007}.  The distribution of the non-affine displacements
$d\tilde{x}$ is also included in Fig.~\ref{fig:20}(a),
and coincides with $P(dy)$.

We now turn to faster shear, $\dot\gamma=0.019$~s$^{-1}$. We first consider a 2D image series taken at
$z=18$~$\mu$m from the 3D stacks. From correlation we find a uniform motion with a constant velocity
$V=1.6$~$\mu$m/frame. To compare with Sec.~\ref{subsubsec:unilimit}, using $\Delta \sim 1.8$~$\mu$m as the
particle spacing, this corresponds to a reduced shift $s/\Delta \simeq 1$ between frames. We locate the particles
in 2D, and track them both with iterated CG tracking, and with correlated image tracking, using
Eq.~\ref{eq:general-advection-removal}. In Fig.~\ref{fig:21} the frame to frame displacement distributions from iterated
tracking are shown, with $P(dx)$ evaluated in a co-moving frame so that $\langle dx \rangle=0$. Thse distributions are
cut-off at $dr=R_T=1$~$\mu$m by definition; the tracking program does not consider possible displacements larger
than $R_T$ (frame-to-frame). $P_{IT}(dx)$ is asymmetric, similar to the MC data in Fig.~\ref{fig:10}(b).

\begin{figure}
\includegraphics[width=0.45\textwidth,clip]{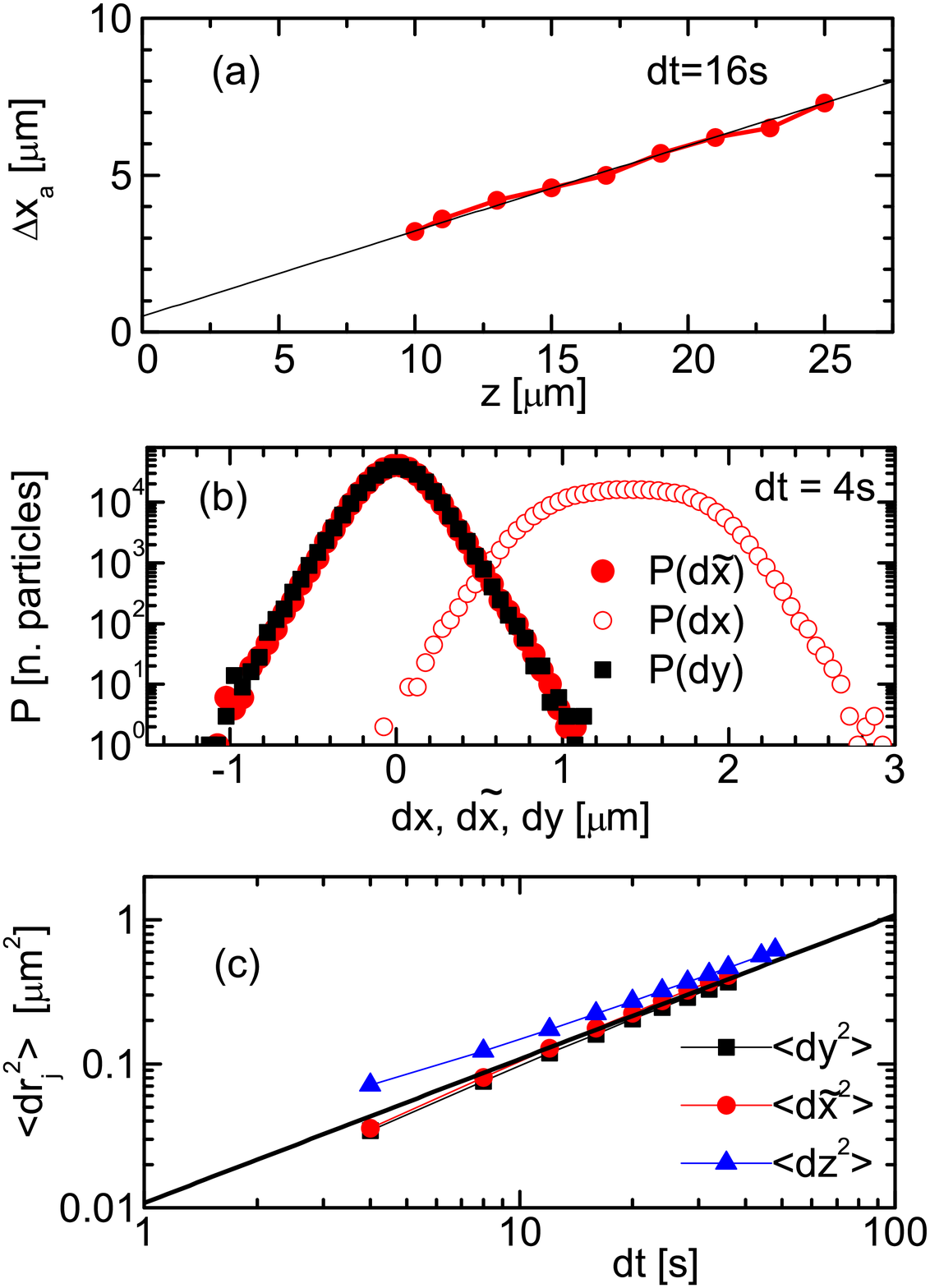}
\caption{3D analysis of a sheared glass at $\dot\gamma=0.019$~s$^{-1}$. (a) (\Red{$\bullet$}) accumulated
displacement $\Delta x_a(z_r,t,dt)$ from image correlation, for $t=40$~s over $dt=16$~s ($4$ frames). Connecting
lines represent the interpolating profile, defining $\Delta x(z,t,dt)$. Line: linear fit giving an accumulated
strain $d\Delta x_a/dz=0.28$. (b) Distribution of frame to frame displacements $P(dx)$ and $P(dy)$ after tracking
in the CM frame ($R_T=1.1$~$\mu$m) and restoring the coordinates in the laboratory frame. Also shown is
$P(d\tilde{x})$ of the non-affine $x$-displacements, using Eq.~\ref{eq:nonaffinedx} and
$\dot\gamma=0.019$~s$^{-1}$. (c) the (non-affine) MSD in the three directions. The data are consistent with those
in Fig.~\ref{fig:21}, inset. Line: $\langle dy^2(t) \rangle=2Dt$ with $D=5.4 \times 10^{-3}$~$\mu{\rm m}^2$/s.}
\label{fig:22}
\end{figure}

This changes when we use correlated image tracking and then examine the measured displacements in the co-moving
frame. Now, the displacement distribution functions, Fig.~\ref{fig:21}, are no longer cut-off and coincide for
$dy$ and $dx$ (in the co-moving frame), indicating correct tracking. The inset shows the resulting MSDs $\langle
dy^2(t) \rangle$ and $\langle dx^2(t) \rangle$ (the latter again in the co-moving frame). As for slow shear, the
dynamics is nearly isotropic. A more detailed description on the shear-induced structural relaxation is given in
\cite{BesselingPRL2007}. From the frame to frame MSD $\langle dy^2(t=4$~s$) \rangle \simeq 0.05$~$\mu{\rm m}^2$ we
obtain $\langle dr^2 \rangle/\Delta^2=\epsilon^2 \simeq 0.03$, within the limits for tracking in a concentrated
quiescent system, Fig.~\ref{fig:8}.

To analyze the 3D data, we use correlated image tracking where the correlation procedure is performed on image
sequences at different heights $z_r=10+2r$~$\mu$m ($r$ integer), as shown in Fig.~\ref{fig:13}(a).
From this, we obtain the accumulated displacements profile $\Delta x_a(z_r,t,dt)=\sum_{t-dt}^{t}\Delta x(z_r,t)$,
an example of which is shown in Fig.~\ref{fig:22}(a) for $t=40$~s and $dt=16$~s. The profile is linear and
again shows approximately stick boundary conditions. Time averages $\langle \Delta x_a(z_r,t,dt=16$~s$)
\rangle_{t}$ (not shown) virtually overlap these data, showing that the flow is steady. To subtract this advected
motion from the particle coordinates, we use the linear interpolation profile $\Delta x(z,t,dt)$ shown by the
lines connecting the symbols. Since both accumulated strain and non-affine displacements are large in this case,
we performed both the subtraction of $\Delta x(z,t)$ (see Eq.~\ref{eq:general-advection-removal}) and the
``piecewise tracking'', in intervals of $10$ frames ($\Delta \gamma= 80\%$), as described in
Sec.~\ref{subsec:limitcorrection}. The resulting displacement distribution functions are shown in Fig.~\ref{fig:22}(b), both for the
non-affine and the real displacements. Note the large range, $0$~$\mu$m~$\leq dx \leq 3$~$\mu$m, of the latter,
resulting from the strong $z$ gradient in advected motion. Finally, we show the MSDs for $\tilde{x}$, $y$ and $z$
calculated from these 3D data. The results for $d\tilde{x}^2$ and $dy^2$ match those in the inset to
Fig.~\ref{fig:21} while $dz^2$ shows that despite the large shear rate, the dynamics remain nearly
isotropic.

\subsection{Rheology and velocity profiling}
\label{subsec:rheoscopeapp}

As a last example we describe the results of experiments on a more dilute colloidal suspension using the confocal
rheoscope in cone-plate geometry. Here we measure simultaneously the rheological response and map the velocity
profile during flow, Figs.~\ref{fig:6},\ref{fig:7}.

\begin{figure}
\includegraphics[width=0.45\textwidth,clip]{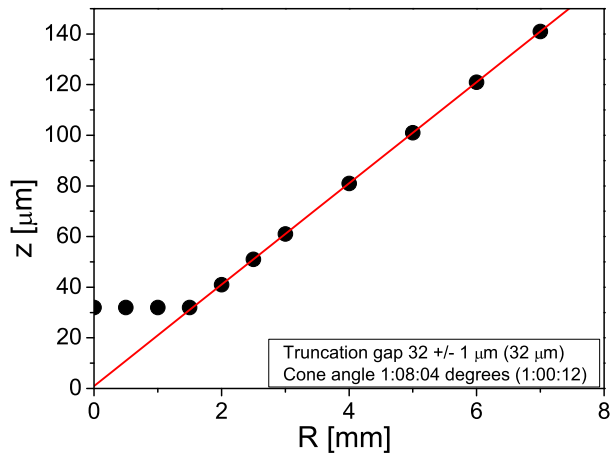} \caption{Gap profile of the cone-plate geometry, measured by confocal microscopy with fluorescent
coating on both the bottom (glass) surface and the surface of the cone (radius $20$~mm). The truncation gap is
clearly visible and the profile extrapolates at zero height in the centre. The nominal values of the truncation
gap and of the cone angle are in brackets in the figure legend.} \label{fig:23}
\end{figure}

The sample consists of a $\phi \sim 55\%$ suspension of non-fluorescent PMMA-PHS colloids (radius = $150$~nm) in
an index matching (decalin-tetralin) mixture, seeded with $\sim 0.5 \%$ fluorescent tracers (radius = $652$~nm).
Both the cone and the cover slide are coated with a layer of tracer particles. From the image series of the cone
motion, we can verify the rotation speed of the rheometer during operation and by focusing on the top and bottom
coatings we can map the spatial profile of the cone-plate geometry. In Fig.~\ref{fig:23} we show the variation of
the gap size as function of the distance $r$ from the center of the cone measured using the lateral objective
translation. The truncation gap is nicely resolved and the data show that bending of the cover slide is
negligible.

The velocity measurements were performed in two ways. One is the `time-resolved' mode: similarly to what is
described in Sec.~\ref{subsec:3Dtrack}, we rapidly scan 3D stacks of $\sim 20$ to $50$ slices, covering the entire
gap from $z=0$ to $z=\theta r$ with $\theta$ the cone angle, Fig.~\ref{fig:7}(a). We use an oil-immersion
objective with $200$~$\mu$m working distance ($60\times$ magnification) and scan at equal speed both up and down
to avoid disturbance of the sample due to large sudden displacements of the objective. By extracting the image
series at each height and using the correlation method (Sec.~\ref{subsubsec:citdescription}), we obtain the time
resolved displacements and shear profile. Typically it takes $\Delta t \sim 1$~s to scan the gap. For the
particular case of images containing tracers, displacements up to approximately half the image size $S$ can be
measured. Therefore, this mode is successful for maximum velocities $v \sim S/(2 \Delta t) \sim 50$~$\mu{\rm
m}/$s, with $S=100$~$\mu$m, which translates to $\dot \gamma_a r < S/(2 \theta \Delta t)$. In the second mode
(`stepping') we simply record a time series at each $z$ and reconstruct the velocity profile $v(z)$. Here the
maximum velocity which can be measured is considerably larger: $v \sim S f_{\rm scan}/2 \sim 5$~mm/s for $f_{\rm
scan}=100$~s$^{-1}$. This corresponds to a maximum shear rate of $\dot\gamma \sim 30$~s$^{-1}$ for $r=10$~mm. Note
that even larger velocities can be measured by using smaller magnification objectives.

\begin{figure}
\includegraphics[width=0.45\textwidth,clip]{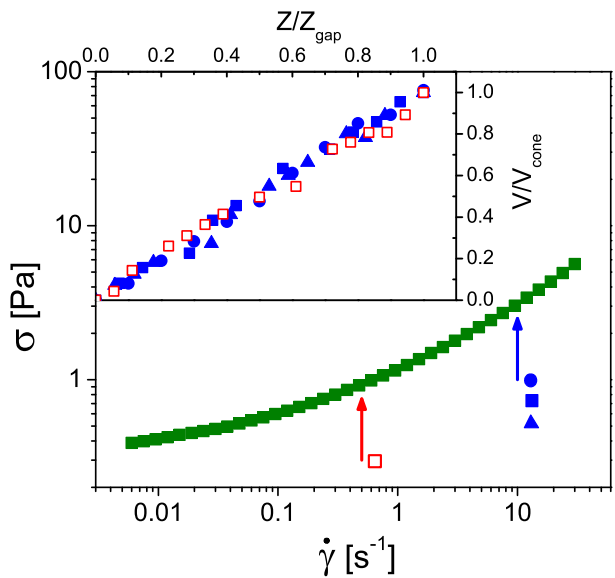}
\caption{Main panel: steady state stress $\sigma$ versus applied shear rate $\dot\gamma$ for a $\phi \sim 55 \%$
suspension of hard-sphere colloids of radius $150$~nm. Inset: velocity profiles, normalized by the velocity of the
cone $V_{cone}$, as a function of the height, normalized by the local gap size $Z_{gap}=\theta r$ of the geometry.
Data for $\dot \gamma=10$~s$^{-1}$ are taken at different positions $r=2.5$~mm (\Blue{$\bullet$}), $r=4$~mm
(\Blue{$\blacksquare$}) and $r=5$~mm (\Blue{$\blacktriangle$}) and for $\dot\gamma=0.5$~s$^{-1}$ data are taken at
$r=5$~mm (\Red{$\square$}).} \label{fig:24}
\end{figure}

Figure~\ref{fig:24} (main panel) shows the steady state flow curve of the sample. Since the suspension at this
volume fraction is in the proximity of the glass transition ($\phi_g \sim 57\%$), the expected low-rate Newtonian
regime occurs at shear rates below our experimental window. In the regime we measured, $0.006$~s$^{-1} <
\dot\gamma< 30$~s$^{-1}$ the sample exhibits strongly nonlinear rheology with pronounced shear thinning response
(the viscosity decreases dramatically on increasing shear rate). In the inset we show some velocity profiles,
measured in the `stepping' mode at different shear rates and various distances $r$ from the center of the cone.
The corresponding gap sizes range from $Z_{gap} \simeq 50$~$\mu$m at $r=2.5$~mm to $Z_{gap} \simeq 100$~$\mu$m at
$r=5$~mm. It is clear that the normalized profiles are linear, independent of $\dot \gamma$ and $r$. In the case
of larger volume fractions, above the glass transition, $\phi > \phi_g$, we observe either slip or shear
localization, depending on the boundary conditions. These results will be discussed in detail elsewhere
\cite{Ballesta2008,BesselingTBP}.

\section{Conclusion}
\label{sec:conclusion}

In this paper we have described new instrumentation and analysis algorithms for 2D and 3D imaging studies of
concentrated (colloidal) suspensions during flow. The combination of fast confocal microscopy and controlled flow,
such as in a rheometer where simultaneous rheological information is available, opens up new horizons for the
study of driven soft matter systems at high concentrations. Our evaluation of the CG tracking algorithm
establishes the validity of previous experiments on colloidal dynamics. Finally, our method for particle tracking
in a locally co-moving frame allows us to investigate the affine and non-affine dynamics of colloids during flow
up to relatively large velocity (gradients), limited primarily by the amount of non-affine motion during flow. The
method could therefore be of use in a variety of other applications, including the study of granular flow at
single-particle level.

\section{Acknowledgements}

We thank K.N. Pham, J. Arlt and N. Pham for advice on the experiments, A.B. Schofield for
particle synthesis, A. Garrie, A. Downie, and V. Devine for technical support, and E. Kim for providing the 3D
Monte Carlo simulation data for hard-spheres. R. Besseling and W.C.K. Poon acknowledge funding through EPSRC
GR/S10377/01 and EP/D067650. L. Isa was funded by the EU network MRTN-CT-2003-504712. E.R. Weeks was funded
through NSF DMR-0603055 (US).


\begin{thebibliography}{00}


\bibitem{PuseyReview91}
P.~N. Pusey, in: J.~P. Hansen, D.~Levesque, J.~Zinn-Justin (Eds.), Liquids,
  Freezing and Glass Transition, Elsevier, 1991, Ch.~10, pp. 767--942.

\bibitem{PoonReview95}
W.~C.~K. Poon, P.~N. Pusey, in: M.~Baus, L.~F. Full, J.~P. Ryckaert (Eds.),
  Observation, Prediction and Simulation of Phase Transitions in Complex
  Fluids, Kluwer, 1995, Ch.~1, pp. 3--51.

\bibitem{poon2}
W.~C.~K. Poon, The physics of a model colloid--polymer mixture, J.
  Phys.-Condens. Mat. 14 (2002) R859--R880.

\bibitem{Aarts04}
D.~G. A.~L. Aarts, M.~Schmidt, H.~N.~W. Lekkerkerker, Interfacial dynamics in
  demixing systems with ultralow interfacial tension, Science 304~(5672) (2004)
  847--850.

\bibitem{AndersonReview02}
V.~J. Anderson, H.~N.~W. Lekkerkerker, Insights into phase transition kinetics
  from colloid science, Nature 416~(6883) (2002) 811--815.

\bibitem{Dawson2002}
K.~A. Dawson, The glass paradigm for colloidal glasses, gels, and other
  arrested states driven by attractive interactions, Curr. Opin. Colloid
  Interface Sci. 7 (2002) 218--227.

\bibitem{Sciortino05}
F.~Sciortino, P.~Tartaglia, Glassy colloidal systems, Adv. Phys. 54~(6-7)
  (2005) 471--524.

\bibitem{Zaccarelli07}
E.~Zaccarelli, Colloidal gels: equilibrium and non-equilibrium routes, J.
  Phys.-Condens. Mat. 19 (2007) 323101.

\bibitem{Fielding00}
S.~M. Fielding, P.~Sollich, M.~E. Cates, Aging and rheology in soft materials,
  J. Rheol. 44 (2000) 323--369.

\bibitem{FuchsCates02}
M.~Fuchs, M.~E. Cates, Theory of nonlinear rheology and yielding of dense
  colloidal suspensions, Phys. Rev. Lett. 89 (2002) 248304.

\bibitem{Schweizer05}
V.~Kobelev, K.~S. Schweizer, Strain softening, yielding, and shear thinning in
  glassy colloidal suspensions, Phys. Rev. E 71 (2005) 021401.

\bibitem{PasteReview06}
D.~I. Wilson, S.~L. Rough, Exploiting the curious characteristics of dense
  solid-liquid pastes, Chem. Eng. Sci. 61 (2006) 4147--4154.

\bibitem{LewisReview00}
J.~A. Lewis, Colloidal processing of ceramics, J. Am. Ceramic Soc. 83 (2000)
  2341--2359.

\bibitem{Jaeger1996}
H.~M. Jaeger, S.~R. Nagel, R.~P. Behringer, Granular solids, liquids, and
  gases, Rev. Mod. Phys. 68~(4) (1996) 1259--1273.

\bibitem{isa2}
L.~Isa, R.~Besseling, W.~C.~K.~Poon, Shear zones and wall slip in the capillary flow of concentrated
  colloidal suspensions, Phys. Rev. Lett. 98~(18) (2007) 198305.

\bibitem{Coussot99}
P.~Coussot, C.~Ancey, Rheophysical classification of concentrated suspensions
  and granular pastes, Phys. Rev. E 59 (1999) 4445--4457.

\bibitem{Bombannes}
P.~Lindner, T.~Zemb (Eds.), Neutron X-rays \& light: scattering methods applied
  to soft condensed matter, North Holland, 2002.

\bibitem{PerrinAtoms}
J.~B. Perrin, Atoms, Constable, London, 1916.

\bibitem{weeksrev2}
V.~Prasad, D.~Semwogerere, E.~R. Weeks, Confocal microscopy of colloids, J.
  Phys.: Condens. Mat. 19 (2007) 113102.

\bibitem{CrockerGrierJColIntSc96_tracking}
J.~C. Crocker, D.~G. Grier, Methods of digital video microscopy for colloidal
  studies, J. Colloid Interface Sci. 179 (1996) 298.

\bibitem{Maret00}
K.~Zahn, G.~Maret, Dynamic criteria for melting in two dimensions, Phys. Rev.
  Lett. 85 (2000) 3656--3659.

\bibitem{Maret03}
K.~Zahn, A.~Wille, G.~Maret, S.~Sengupta, P.~Nielaba, Elastic properties of 2d
  colloidal crystals from video microscopy, Phys. Rev. Lett. 90 (2003) 155506.

\bibitem{Maret08}
P.~Ebert, P.~Keim, G.~Maret, Local crystalline order in a 2d colloidal glass
  former, Eur. Phys. J. E 26 (2008) 161--168.

\bibitem{Elliot01}
M.~S. Elliot, W.~C.~K. Poon, Conventional optical microscopy of colloidal
  suspensions, Adv. Colloid Interface Sci. 92 (2001) 133--194.

\bibitem{Bristol97}
M.~S. Elliot, B.~T.~F. Bristol, W.~C.~K. Poon, Direct measurement of stacking
  disorder in hard sphere colloidal crystals, Physica A 235 (1997) 216--223.

\bibitem{wilson}
T.~Wilson, Confocal Microscopy, Academic Press, San Diego, 1990.

\bibitem{vanblaaderen1}
A.~van Blaaderen, P.~Wiltzius, Real-space structure of colloidal hard-sphere
  glasses, Science 270~(5239) (1995) 1177--1179.

\bibitem{dinsmore}
A.~D. Dinsmore, E.~R.~Weeks, V.~Prasad, A.~C.~Levitt, D.~A.~Weitz, Three--dimensional confocal microscopy of
colloids,
  App. Optics 40 (2001) 4152--4159.

\bibitem{habdas02}
P.~Habdas, E.~R. Weeks, Video microscopy of colloidal suspensions and colloidal
  crystals, Curr. Opin. Colloid In. 7~(3-4) (2002) 196--203.

\bibitem{weeksrev1}
D.~Semwogerere, E.~R. Weeks, Confocal microscopy, Encyclopedia of Biomaterials
  and Biomedical Enigineering.

\bibitem{Weitz01}
U.~Gasser, E.~R. Weeks, A.~Schofield, P.~N. Pusey, D.~A. Weitz, Real-space
  imaging of nucleation and growth in colloidal crystallization, Science 292
  (2001) 258--262.

\bibitem{kegel}
W.~K. Kegel, A.~van Blaaderen, Direct observation of dynamical heterogeneities
  in colloidal hard-spheres suspensions, Science 287 (2000) 290--293.

\bibitem{weeks1}
E.~R. Weeks, J.~C. Crocker, A.~C. Levitt, A.~Schofield, D.~A. Weitz,
  Three--dimensional direct imaging of structural relaxation near the colloidal
  glass transition, Science 287 (2000) 627--631.

\bibitem{PoonReview09}
W.~C.~K. Poon, R.~Besseling, L.~Isa, A.~B. Schofield, Imaging of hard-sphere
  colloids under flow, Adv. Polymer Sci. to be published.

\bibitem{pham3}
K.~N. Pham, G.~Petekidis, D.~Vlassopoulos, S.~U. Egelhaaf, P.~N. Pusey,
  W.~C.~K. Poon, Yielding of colloidal glasses, Europhys. Lett. 75 (2006)
  624--630.

\bibitem{Pham08}
K.~N. Pham, G.~Petekidis, D.~Vlassopoulos, S.~U. Egelhaaf, W.~C.~K. Poon, P.~N.
  Pusey, Yielding behavior of repulsion- and attraction-dominated colloidal
  glasses, J. Rheol. 52 (2008) 649--676.

\bibitem{RusselGrantColSurfA2000_slipyield}
W.~Russel, M.~Grant, Distinguishing between dynamic yielding and wall slip in a
  weakly flocculated colloidal dispersion, Colloids Surf. A 161 (2000) 271.

\bibitem{BuscalJRheo93_slip}
R.~Buscall, J.~I. McGowan, A.~J. Morton-Jones, The rheology of concentrated
  dispersions of weakly attracting colloidal particles with and without wall
  slip, Colloids Surf. A 37 (1993) 621.

\bibitem{Olmsted08}
P.~D. Olmsted, Perspectives on shear banding in complex fluids, Rheol. Acta 47
  (2008) 283--300.

\bibitem{PIV}
M.~Raffel, C.~Willert, J.~Kompenhans, Particle Image Velocimetry, a Practical
  Guide, Springer, Berlin, 1998.

\bibitem{MeekerPRL04_rheoslip}
S.~P. Meeker, R.~T. Bonnecaze, M.~Cloitre, Slip and flow in soft particle
  pastes, Phys. Rev. Lett. 92~(19) (2004) 198302.

\bibitem{MeekerJRheo2004_slipandflowrheo}
S.~P. Meeker, R.~T. Bonnecaze, M.~Cloitre, Slip and flow in pastes of soft
  particles: Direct observation and rheology, J. Rheol. 48~(6) (2004)
  1295--1320.

\bibitem{SalmonEPJAP2003_heterodyneDLS}
J.-B. Salmon, S.~Manneville, A.~Colin, B.~Pouligny, An optical fiber based
  interferometer to measure velocity profiles in sheared complex fluids, Eur.
  Phys. J. Appl. Phys. 22 (2003) 143.

\bibitem{MannevilleEPJAP04_ultrasound}
S.~Manneville, L.~B\'{e}cu, A.~Colin, High-frequency ultrasonic speckle
  velocimetry in sheared complex fluids, Eur. Phys. J. Appl. Phys. 28 (2004)
  361.

\bibitem{BecuPRL06_twoemulsions}
L.~B\'{e}cu, S.~Manneville, A.~Colin, Yielding and flow in adhesive and
  nonadhesive concentrated emulsions, Phys. Rev. Lett. 96~(13) (2006) 138302.

\bibitem{BecuPRE07_rheomic}
L.~B\'{e}cu, D.~Anache, S.~Manneville, A.~Colin, Evidence for three-dimensional
  unstable flows in shear-banding wormlike micelles, Phys. Rev. E 76~(1) (2007)
  011503.

\bibitem{fukushima_NMRAnnRevFlMech99}
E.~Fukushima, Nuclear magnetic resonance as a tool to study flow, Annu. Rev.
  Fluid Mech. 31 (1999, and references therein) 95--123.

\bibitem{CallaghanRepProgPhys1999_rheoNMR}
P.~T. Callaghan, Rheo-nmr: nuclear magnetic resonance and the rheology of
  complex fluids, Rep. Progr. Phys. 62~(4) (1999) 599--670.

\bibitem{BonnAnnRevFlMech2008_NMR}
D.~Bonn, S.~Rodts, M.~Groenink, S.~Rafa{\"{i}}, N.~Shahidzadeh-Bonn,
  P.~Coussot, Some applications of magnetic resonance imaging in fluid
  mechanics: Complex flows and complex fluids, Annu. Rev. Fluid Mech. 40 (2008)
  209--233.

\bibitem{GladdenMeasSciTech96_NMR}
L.~F. Gladden, P.~Alexander, Applications of nuclear magnetic resonance imaging
  in process engineering, Meas. Sci. Technol. 7~(3) (1996) 423--435.

\bibitem{ovarlez2}
G.~Ovarlez, F.~Bertrand, S.~Rodts, Local determination of the constitutive law
  of a dense suspension of noncolloidal particles through magnetic resonance
  imaging, J. Rheol. 50~(3) (2006) 259--292.

\bibitem{RaynaudCoussotJRheo02_NMRyield}
J.~S. Raynaud, P.~Moucheront, J.~C. Baudez, F.~Bertrand, J.~P. Guilbaud,
  P.~Coussot, Direct determination by nuclear magnetic resonance of the
  thixotropic and yielding behavior of suspensions, J. Rheol. 46~(3) (2002)
  709--732.

\bibitem{HuangBonnPRL04_wetgranularflow}
N.~Huang, G.~Ovarlez, F.~Bertrand, S.~Rodts, P.~Coussot, D.~Bonn, Flow of wet
  granular materials, Phys. Rev. Lett. 94 (2005) 028301.

\bibitem{fall}
A.~Fall, N.~Huang, F.~Bertrand, G.~Ovarlez, D.~Bonn, Shear thickening of
  cornstarch suspensions as a reentrant jamming transition, Phys. Rev. Lett.
  1~(1) (2008) 018301.

\bibitem{BreedveldJFlMech98}
V.~Breedveld, D.~van~den Ende, A.~Tripathi, A.~Acrivos, The measurement of the
  shear-induced particle and fluid tracer diffusivities in concentrated
  suspensions by a novel method, J. Fl. Mech. 375 (1998) 297.

\bibitem{BreedveldPRE2001_nonbrowniantracerdiffusion}
V.~Breedveld, D.~van~den Ende, M.~Bosscher, R.~J.~J.~Jongschaap, J.~Mellema, , Measuring shear-induced
  self-diffusion in a counterrotating geometry, Phys. Rev. E 63 (2001) 021403.

\bibitem{haw2}
M.~D. Haw, W.~C.~K. Poon, P.~N. Pusey, Direct observation of
  oscillatory-shear-induced order in colloidal suspensions, Phys. Rev. E 57~(6)
  (1998) 6859--6864.

\bibitem{haw3}
M.~D. Haw, W.~C.~K. Poon, P.~N. Pusey, P.~Hebraud, F.~Lequeux, Colloidal
  glasses under shear strain, Phys. Rev. E 58~(4) (1998) 4673--4682.

\bibitem{SmithPoonPRE2007}
P.~Smith, G.~Petekidis, S.~U. Egelhaaf, W.~C.~K. Poon, Yielding and
  crystallization of colloidal gels under oscillatory shear, Phys. Rev. E 76
  (2007) 041402.

\bibitem{BiehlPalbergEPL04}
R.~Biehl, T.~Palberg, Modes of motion in a confined colloidal suspension under
  shear, Europhys. Lett. 66~(2) (2004) 291--295.

\bibitem{tolpekin}
V.~A. Tolpekin, M.~H.~G. Duits, D.~van~den Ende, J.~Mellema, Aggregation and
  breakup of colloidal particle aggregates in shear flow, studied with video
  microscopy, Langmuir 20~(7) (2004) 2614--2627.

\bibitem{hoekstra}
H.~Hoekstra, J.~Vermant, M.~J, Flow-induced anisotropy and reversible
  aggregation in two-dimensional suspensions, Langmuir 21~(24) (2003)
  9134--9141.

\bibitem{stancik}
E.~J. Stancik, G.~T. Gavranovic, M.~J.~O. Widenbrant, A.~T. Laschtisch,
  J.~Vermant, G.~G. Fuller, Structure and dynamics of particle monolayers at a
  liquid-liquid interface subject to shear flow, Faraday Discuss. 123 (2003)
  145--156.

\bibitem{VaradanSolomonJRheo2003_gelflowconfocal}
P.~Varadan, M.~J. Solomon, Direct visualization of flow-induced microstructure
  in dense colloidal gels by confocal laser scanning microscopy, J. Rheol.
  47~(4) (2003) 943--968.

\bibitem{CohenWeitzPRL04_confinedshear}
I.~Cohen, T.~Mason, D.~Weitz, Shear-induced configurations of confined
  colloidal suspensions, Phys. Rev. Lett. 93 (2004) 046001.

\bibitem{SolomonJCP06_shearcrystalCF}
T.~Solomon, M.~J. Solomon, Stacking fault structure in shear-induced colloidal
  crystallization, J. Chem. Phys. 124~(13) (2006) 134905.

\bibitem{derks}
D.~Derks, H.~Wisman, A.~van Blaaderen, A.~Imhof, Confocal microscopy of
  colloidal dispersions in shear flow using a counter-rotating cone-plate shear
  cell, J. Phys.-Condens. Mat. 16 (2004) S3917--S3927.

\bibitem{wu2}
Y.~L. Wu, J.~H.~J. Brand, J.~L.~A. van Gemert, J.~Verkerk, H.~Wisman, A.~van
  Blaaderen, A.~Imhof, A new parallel plate shear cell for {\textit{in situ}}
  real-space measurements of complex fluids under shear flow, Rev. Sci.
  Instrum. 78 (2007) 103902.

\bibitem{BesselingPRL2007}
R.~Besseling, E.~R. Weeks, A.~B. Schofield, W.~C.~K. Poon, Three-dimensional
  imaging of colloidal glasses under steady shear, Phys. Rev. Lett. 99 (2007)
  028301.

\bibitem{XuReeveseRSciInstr04_trackingerrors}
H.~Xu, A.~P. Reeves, M.~Y. Louge, Measurement error in the mean and fluctuation
  velocities of spherical grains form a computer analysis of digital images,
  Rev. Sci. Instrum. 75~(4) (2004) 811--819.

\bibitem{antl}
L.~Antl, J.~W. Goodwin, R.~D. Hill, R.~H. Ottewill, S.~M. Owens, S.~Papworth,
  J.~A. Waters, The preparation of poly(methyl methacrylate) lattices in
  non-aqueous media, Colloid. Surface. 17~(1) (1986) 67--78.

\bibitem{ackerson}
B.~J. Ackerson, P.~N. Pusey, Shear--induced order in suspensions of
  hard--spheres, Phys. Rev. Lett. 61~(8) (1988) 1033--1036.

\bibitem{yethiraj}
A.~Yethiraj, A.~{van~Blaaderen}, A colloidal model system with an interaction
  tunable from hard sphere to soft and dipolar, Nature 421~(6922) (2003)
  513--517.

\bibitem{lu2}
P.~J. Lu, J.~C. Conrad, H.~M. Wyss, A.~B. Schofield, D.~A. Weitz, Fluids of
  clusters in attractive colloids, Phys. Rev. Lett. 96 (2006) 028306.

\bibitem{Ballesta2008}
P.~Ballesta, R.~Besseling, L.~Isa, G.~Petekidis, W.~C.~K.~Poon, arXiv:0807.1437v1 (2008).

\bibitem{isa1}
L.~Isa, R.~Besseling, E.~R.~Weeks, W.~C.~K.~Poon, Experimental studies of the flow of concentrated hard sphere
  suspensions into a constriction, J. Phys.--Conference Series 40 (2006)
  124--132.

\bibitem{idlwebsite}
E.~R. Weeks, J.~C. Crocker, Particle tracking using {IDL}: website,
  http://www.physics.emory.edu/$\sim$weeks/idl/.

\bibitem{BrangwynneKoenderinkBioPhysJ2007_filamenttracking}
C.~P. Brangwynne, G.~H. Koenderink, E.~Barry, Z.~Dogic, F.~C. MacKintosh, D.~A.
  Weitz, Bending dynamics of fluctuating biopolymers probed by automated
  high-resolution filament tracking, Biophys. J. 93 (2007) 346--359.

\bibitem{JenkinsEgelhaafAdvColIntSci08}
M.~Jenkins, S.~U. Egelhaaf, Confocal microscopy of colloidal particles: Towards
  reliable, optimum coordinates, Adv. Col. Int. Sci. 136~(1--2) (2008) 65--92.

\bibitem{pusey_hydro}
P.~N. Pusey, R.~J.~A. Tough, Hydrodynamic interactions and diffusion in
  concentrated particle suspensions, Faraday Discuss. 76~(76) (1983) 123--136.

\bibitem{BeenhakkerMazurPhys84}
C.~Beenhakker, P.~Mazur, Diffusion of spheres in a concentrated suspension .2.,
  Physica A 126 (1984) 349--370.

\bibitem{TokuyamaPRE94_diffusion}
M.~Tokuyama, I.~Oppenheim, Dynamics of hard-spheres suspensions, Phys. Rev. E.
  50 (1994) R16.

\bibitem{BradyJChemPhys93_rheocol}
J.~Brady, Brownian-motion, hydrodynamics, and the osmotic-pressure, J. Chem.
  Phys. 567 (1993) 99.

\bibitem{MegenPRE98_tracersinglass}
W.~van Megen, T.~C. Mortensen, S.~R. Williams, J.~M{\"{u}}ller, Measurements of
  the self--intermediate scattering function of suspensions of hard spherical
  particles near the glass transition, Phys. Rev. E 58~(5) (1998) 6073--6085.

\bibitem{SteagerAPL07_baterialtrack}
E.~Steager, C.-B. Kim, J.~Patel, S.~Bith, C.~Naik, L.~Reber, M.~J. Kim, Control
  of microfabricated structures powered by flagellated bacteria using
  phototaxis, Appl. Phys. Lett 90 (2007) 263901.

\bibitem{RogersBioPhJ2008_tracking}
S.~S. Rogers, T.~A. Waigh, J.~R. Lu, Intracellular microrheology of motile
  amoeba proteus, Biophys. J. 94~(8) (2008) 3313--3322.

\bibitem{chetverikov01particle}
D.~Chetverikov, Particle image velocimetry by feature tracking, Lecture Notes
  in Computer Science 2124 (2001) 325.

\bibitem{DoliwaPRL98_HSglasssimu_hetero}
B.~Doliwa, A.~Heuer, Cage effect, local anisotropies, and dynamic
  heterogeneities at the glass transition: A computer study of hard spheres,
  Phys. Rev. Lett. 80 (1998) 4915.

\bibitem{habdas}
P.~Habdas, D.~Schaar, A.~C. Levitt, E.~R. Weeks, Forced motion of a probe
  particle near the colloidal glass transition, Europhys. Lett. 67~(3) (2004)
  477--483.

\bibitem{You05}
R.~R. Huilgol, Z.~You, Application of the augmented {L}agrangian method to
  steady pipe flows of {B}ingham, {C}asson and {H}erschel--{B}ulkley fluids, J.
  Non-Newton. Fluid Mech. 128 (2005) 126--143.

\bibitem{frank03}
M.~Frank, D.~Anderson, E.~R. Weeks, J.~F. Morris, Particle migration in
  pressure-driven flow of a brownian suspension, J. Fluid Mech 493 (2003)
  363--378.

\bibitem{BesselingTBP}
R.~Besseling, P.~Ballesta, L.~Isa, G.~Petekidis, W.~C.~K.~Poon, Shear localization in
  hard-sphere colloidal glasses, in preparation.

\bibitem{YamamotoOnukiPRE98_SCLiq_rheodif}
R.~Yamamoto, A.~Onuki, Dynamics of highly supercooled liquids: Heterogeneity,
  rheology, and diffusion, Phys. Rev. E 58~(3) (1998) 3515--3529.

\bibitem{Jaster}
A.~Jaster, Orientational order of the two-dimensional hard-disk system,
  Europhys. Lett. 42~(3) (1998) 277--281.

\end{thebibliography}
\end{document}